%% file: usenix.tex
\documentclass[letterpaper,twocolumn,10pt]{article}

\newif\ifnotnewacm
\notnewacmtrue
\newif\ifheadnice
\headnicefalse

\usepackage{usenix}

\usepackage{xspace}
\usepackage{xcolor}

\newcommand{\secref}[1]{\S\ref{#1}}
\newcommand{\figref}[1]{Fig.~\ref{#1}}
\newcommand{\tabref}[1]{Tab.~\ref{#1}}
\newcommand{\eqnref}[1]{Eqn.~\ref{#1}}
\newcommand{\algoref}[1]{Alg.~\ref{#1}}

\def\sys{\textsc{Blade}\xspace}

\renewcommand{\paragraph}[1]{\vspace{2pt plus 0pt minus 2pt}\noindent{\bfseries #1}}

\input{preamble}

\newcommand{\affmark}[1]{\textsuperscript{#1}}

\makeatletter
\def\@maketitle{%
  \newpage
  \null
  \vskip 1.6em
  \begin{center}%
    {\Large\bfseries \@title \par}%
    \vskip 1.0em
    {\normalsize
      \lineskip 0.5em
      \begin{tabular}[t]{c}\@author\end{tabular}\par
    }%
  \end{center}%
  \vskip 1.8em % 标题/作者块后留白：这里增大 -> Abstract+正文整体下移
}
\makeatother

\begin{document}
\pagestyle{empty}
\date{}

%\title{\sys: Adaptive Wi-Fi Contention Control for Next-Generation Real-Time Communication}
\title{\sys: Adaptive Wi-Fi Contention Control \\for Next-Generation Real-Time Communication}
% =========================================================
% 作者与单位：字号完全一致（不要 \large / \small）
% 风格如果你想“像 NSDI 示例”更统一：这里用 \itshape（作者与单位同风格）
% 不写通信作者图例：不再输出 “Corresponding authors”
% =========================================================
\author{%
\normalfont
Fengqian Guo\affmark{1}\thanks{equal contribution}\quad
Yuhan Zhou\affmark{2}\affmark{1}\footnotemark[1]\quad
Longwei Jiang\affmark{1}\quad
Congcong Miao\affmark{1}\quad
Yuxin Liu\affmark{3}\\
Chenren Xu\affmark{2}\quad
Hancheng Lu\affmark{4}\quad
Chang Wen Chen\affmark{5}\quad
Yaxiong Xie\affmark{3}\thanks{Corresponding authors.}\quad
Honghao Liu\affmark{1}\footnotemark[2]\\[0.8ex]
\begin{minipage}{0.96\linewidth}
\centering\normalfont\itshape
\affmark{1}Tencent \quad
\affmark{2}Peking University \quad
\affmark{3}University at Buffalo, SUNY\\
\affmark{4}Institute of Artificial Intelligence, China \quad
\affmark{5}The Hong Kong Polytechnic University \quad
\end{minipage}%
}

\maketitle

\input{sections/00-abstract}

\input{sections/index}

\section*{Acknowledgments}
We thank our shepherd, Robert Ricci, and the anonymous reviewers for their valuable comments. We would like to express our sincere gratitude to our colleagues at Tencent START, including Weiting Xiao, Zhenxing Wen, Jianjun Xiao, Nian Wen and Jiafeng Chen, for their invaluable technical support, insightful discussions.%This work was supported in part by the National Science Foundation of China under Grant U21A20452, by the Hong Kong Research Grants Council (GRF-15213322, GRF-15229423), and by the Fundamental Research Funds for the Central Universities (No.\ WK2100250067).

\newpage
\bibliographystyle{unsrt}
\bibliography{ref}

\newpage
\input{sections/99-apendix}

\end{document}

%% file: preamble.tex
\begin{CJK}{UTF8}{gbsn}

\newtheorem{algorithm}{算法}
\renewcommand{\abstractname}{摘要}  % 将Abstract改为摘要
\title{***}
\author{***footnote{电子邮件:**}\\[2ex]
*** \\[2ex]
}
\date{20XX年X月}

\maketitle

\tableofcontents
\newpage
在此输入正文，中英文均可。

\songti 
\fangsong 
\biaosong
\heiti
\kaishu

\end{CJK}

%% file: sections/00-abstract.tex
\begin{abstract}
Next-generation real-time communication (NGRTC) applications, such as cloud gaming and XR, demand consistently ultra-low latency. However, through our first large-scale measurement, we find that despite the deployment of edge servers, dedicated congestion control, and loss recovery mechanisms, cloud gaming users still experience long-tail latency in Wi-Fi networks. 
We further identify that Wi-Fi last-mile access points (APs) serve as the primary latency bottleneck. Specifically, short-term packet delivery droughts, caused by fundamental limitations in Wi-Fi contention control standards, are the root cause. To address this issue, we propose \sys, an adaptive contention control algorithm that dynamically adjusts the contention windows (CW) of all Wi-Fi transmitters based on the channel contention level in a fully distributed manner. 
Our ns3 simulations and real-world evaluations with commercial Wi-Fi APs demonstrate that, compared to standard contention control, \sys reduces Wi-Fi packet transmission tail latency by over $5\times$ under heavy channel contention and significantly stabilizes MAC throughput while ensuring fast and fair convergence. Consequently, \sys reduces the video stall rate in cloud gaming by over 90\%.

\end{abstract}

%% file: sections/index.tex
\input{sections/01-introduction}
\input{sections/20_1_primer}
\input{sections/20_2_motivation}

\input{sections/40-design}

\input{sections/50-implmentation}

\input{sections/60-evaluation}

\input{sections/70-discussion}

\input{sections/80-related}

\input{sections/90-conclusion}

%% file: sections/01-introduction.tex
\section{Introduction} 
%\yuhan{dummy content, ignore this section}
%\yuhan{emphasize online stall rate: 1) add a table of current frame tail latency and video stall rate with wifi; 2) show the influence of stall rate.}
Emerging next generation real time communication (NGRTC) systems
such as cloud gaming \cite{samsumg22, google23-gaming} and Extended Reality (XR) \cite{googlemap23,
youtubevr23} are revolutionizing how users experience interactive digital content.
These applications have been rapidly adopted across both entertainment and business,
with the global cloud gaming market alone growing from \$1,286.6 million 
in 2022 to a projected \$13.6 billion by 2028 \cite{market-report23}.
Such next-generation real-time streaming applications require both \textit{high bandwidth} (\eg \textasciitilde30 Mbps
for cloud gaming \cite{xu22-cloudgaming-mea} and \textasciitilde50 Mbps for XR \cite{mangiante17-vrdemand})
and \textit{consistently low latency} to maintain their interactive nature and deliver immersive user
experiences \cite{shibo24-pudica, yuhan24-augur}.

\noindent\textbf{Long-tail Latency.}
Except average latency and throughput, NGRTC are extremely sensitive to long-tail latency:
even a single latency spike, \textit{i.e.}, latency larger than 200 ms, can trigger a video stall, 
catastrophically disrupting the user’s immersive experience~\cite{yuhan24-augur, zili22-zhuge, zili24-hairpin, shibo24-pudica, zili23-afr, jiangkai2023-zgaming}. 
The stakes are high; prior work has shown that a mere 0.5\%
increase in stall rate can reduce user retention time by a third~\cite{shibo24-pudica, yuhan24-augur}.
Consequently, the viability of this burgeoning multi-billion dollar market
hinges on delivering data with near-perfect punctuality.

To prevent these disruptive stalls, the underlying network must provide highly predictable, timely delivery.
With a video stall defined as any frame delivery taking longer than 200 ms,
NG-RTC applications implicitly demand that the network deliver nearly every frame within this strict
budget; even failures at the 99.99th percentile can severely degrade the user experience.
This imposes a stringent requirement for predictable network performance.

The Internet’s most prevalent last-hop technology, Wi-Fi,
is theoretically incapable of providing this guarantee.
Unlike cellular networks, which leverage centralized scheduling to allocate resources and manage latency~\cite{3gpp-38211, 3gpp-38213},
Wi-Fi’s design is fundamentally distributed and uncoordinated.
Its reliance on a contention-based channel access protocol (CSMA/CA) means that
devices must independently compete for transmission opportunities without any central control.
This lack of coordination makes it impossible to guarantee timely packet delivery,
as latency can vary dramatically based on the instantaneous level of channel contention.

%The central challenge is that the Internet’s most prevalent last-hop technology, Wi-Fi,
%offers no such guarantees.
%
%The problem is a failure of predictability at the wireless edge.
%For an interactive stream,
%%generating video frames every 16.7 ms (at 60 frames/s),
%an end-to-end delivery latency exceeding 200 ms is sufficient to cause a visible stall.
%While other network segments have largely solved this problem,
%\textit{e.g.}, modern congestion control can keep 99th-percentile latency 
%in wired networks below 70 ms~\cite{shibo24-pudica, ray22-sqp, michael23-tambur, yunzhe23-cellfusion, zili24-hairpin},
%and cellular networks leverage centralized scheduling for resource allocation~\cite{3gpp-38211, 3gpp-38213, 3gpp-38913},
%Wi-Fi remains a major source of unpredictable,
%high-tail latency. Its reliance on a distributed,
%contention-based channel access protocol (CSMA/CA) creates inherent variability.
%Given that Wi-Fi accounts for over 80\% of all cloud gaming time,
%addressing its performance limitations is 
%%not just an academic exercise but 
%a critical necessity for the entire NG-RTC ecosystem.

{Our large-scale measurement study on the Tencent START cloud
gaming platform confirms this theoretical limitation (\S\ref{sec:online-measurement}).}
The study, which analyzed 336 million video frames from 200 commercial Wi-Fi access points deployed nationwide,
reveals that while the wired portion of the network (from server to access point) 
maintains low latency, staying below 200 ms even at the 99.99th percentile, 
the total end-to-end latency can exceed 1000ms when the wireless last hop is included.
%the Wi-Fi last hop is the primary source of high tail latency.
This empirical evidence dictates that the solution must reside at the Wi-Fi last hop.
{This is not a problem that traditional end-to-end congestion control can solve, 
as congestion control only mitigates queuing delay rather than reducing the brief,
intermittent sudden jitters inherent to the Wi-Fi last hop.
The catastrophic latency spikes are brief, intermittent, and localized entirely within the Wi-Fi last hop. Therefore, resolving the random jitters of over-the-air transmission is critical for NGRTC.
Our research focus is on the core cause of video stalls: the Wi-Fi last hop.
We address this by designing a deployable Media Access Control (MAC)-layer contention-control algorithm that reduces the tail latency of the
last hop and enhances the smoothness of application-layer transmission.}

\noindent\textbf{Packet-Delivery Drought and its Root Cause.}
Drilling down into the Wi-Fi bottleneck,
our measurements reveal that these latency spikes are caused by a specific,
recurring failure mode we term a \textit{packet-delivery drought}: 
a 200 ms interval during which an access point fails to deliver \textit{a single packet} to a user.
This MAC-layer phenomenon is the direct cause of application-layer failure: video stalls.
Our central empirical finding is that 86.19\% of all video stalls are
directly correlated with the occurrence of at least one such drought,
establishing a near one-to-one mapping between the two.

The root cause of these droughts is not slow physical transmission or a lack of channel capacity.
Our measurements confirm that the time spent on physical packet transmission (PHY TX) is consistently brief,
with a 99.99th-percentile delay below 5 ms.
In stark contrast, the contention interval, 
\textit{i.e.}, the time a device spends waiting for channel access,
exhibits an alarming heavy tail, exceeding 200 ms at the 99.99th percentile.
This delay originates from a fundamental flaw in the IEEE 802.11 CSMA/CA protocol: 
\textit{short-term unfairness driven by its exponential backoff mechanism}.
Specifically, after a collision, a device doubles its contention window (CW),
creating a temporary but severe priority asymmetry.
Other devices with smaller CWs can repeatedly seize the channel while
the device with the large CW is forced to wait, its backoff counter perpetually frozen.
In congested environments,
these interruptions can \textit{extend a simple backoff from milliseconds to hundreds of milliseconds},
starving the device of access and creating a packet-delivery drought.
This is a failure of micro-fairness, not aggregate channel efficiency.

%A natural response might be to leverage existing Wi-Fi Quality of Service (QoS) mechanisms,
%such as the priority queues defined in IEEE 802.11e (EDCA).
%However, this approach is fundamentally flawed in modern, dense environments.
%QoS grants priority by assigning latency-sensitive traffic to queues with smaller,
%more aggressive contention windows.
%While effective for a single high-priority flow,
%this static prioritization backfires when multiple high-throughput NGRTC applications compete simultaneously.
%When several devices all use these privileged, smaller contention windows,
%they do not coordinate; they simply intensify channel contention,
%leading to more frequent collisions and exacerbating the very tail latency issues they are meant to solve.

A conventional approach is to leverage existing Wi-Fi Quality of Service (QoS) mechanisms—such as the priority queues of the Enhanced Distributed Channel Access (EDCA) mechanism defined in the IEEE 802.11e standard. On the one hand, encrypted traffic has become the dominant form of Internet traffic\cite{cloudflare-report}, making the identification of traffic for specific priority levels extremely challenging. On the other hand, in dense network environments, concurrent contention for the channel by multiple high-priority traffic flows merely intensifies channel competition, leading to more frequent packet collisions and thereby exacerbating the tail latency issues that such QoS mechanisms were originally designed to mitigate.
This demonstrates that a simple priority scheme is insufficient.
What is needed is a cooperative mechanism that allows all
devices to adapt to the actual level of channel contention.

\noindent\textbf{Solution: Predictable and Cooperative Contention.}
To eliminate these droughts,
we must replace Wi-Fi’s flawed signaling with a mechanism that enables predictable, cooperative behavior.
We present \sys, an adaptive contention control algorithm that fundamentally changes
how devices perceive and react to network congestion.
The critical flaw in the standard protocol is its reliance on a \textit{local and reactive signal}: a collision.
While all devices listen before talk using clear channel assessment (CCA),
they only adjust their behavior aggressively after their own transmission fails.
This signal is local: uninvolved devices remain oblivious to the contention event;
and reactive: addressing congestion only after it has already caused a failure.
This leads to uncoordinated responses where some devices are forced to wait while others,
with smaller contention windows, continue to seize the channel,
creating the priority asymmetries that cause droughts.

\sys solves this by deriving a universal and proactive signal from the same CCA mechanism.
Instead of waiting for a personal failure,
each device continuously measures the \textit{microscopic access rate} (MAR):
the ratio of successful transmission events (from any device) to the number of idle time slots it observes.
Because the protocol forces every device to defer to any ongoing transmission,
{devices within the same carrier-sense domain typically observe consistent busy/idle slot dynamics}.
    {(Hidden terminals and partial visibility can violate this assumption; we make this explicit and discuss mitigation via RTS/CTS in \secref{sec:mar-def} and \secref{append:hidden-terminal}.)} 
This provides a consistent, shared, and quantitative measure of the channel's current contention level.
By shifting from a local, reactive signal to a universal, proactive one,
\sys enables all devices to act cooperatively.
They adjust their contention windows based on a shared understanding of network congestion,
preventing the short-term unfairness that causes packet-delivery droughts in the first place.
{\sys is a MAC-layer transmitter-side mechanism. In our primary, downlink-dominated cloud gaming setting, deploying \sys on APs (the dominant transmitters) already addresses long-tail contention among neighboring APs and does not require client STA modifications; when uplink traffic is significant, an AP can optionally advertise contention parameters via standards-compliant EDCA parameter sets, or STAs can run \sys locally.}

\sys achieves this cooperative behavior through two core mechanisms.
First, it introduces the universally observable contention signal: MAR.
Second, it employs a \textit{hybrid increase multiplicative decrease} (HIMD) policy that uses MAR as feedback,
enabling co-channel devices to collectively and dynamically adapt
their contention windows to match competition level.
This allows the network to converge on fair and efficient operation without explicit coordination,
proactively preventing the priority asymmetries that cause droughts.

We evaluate \sys through extensive real-world experiments with commercial Wi-Fi APs and ns3 simulations.
The results demonstrate that \sys directly remedies the root cause of tail latency.
Compared to the standard contention control,
\sys reduces Wi-Fi packet transmission tail latency by over 5$\times$ under heavy channel contention.
This MAC-layer improvement translates directly to application-level benefits: for cloud gaming,
{\sys reduces the 99th-percentile video frame delivery latency to \ensuremath{\le}0.5$\times$ the baseline and, consequently,
cuts the video stall rate by over 90\%.

\vspace{1mm}
\noindent\textbf{Contribution}. This paper makes the following contributions:
\begin{itemize}[leftmargin = 0.2cm]
    \setlength{\itemsep}{2pt}
    \vspace{-2mm}
    \item We conduct a large-scale measurement of a commercial cloud gaming
    service and identify that packet-delivery droughts in the Wi-Fi last hop,
    caused by fundamental limitations in standard contention control,
    are the root cause of high tail latency for NGRTC applications.

    \vspace{-2mm}
    \item We design and implement \sys,
    an adaptive contention control algorithm that dynamically and cooperatively adjusts
    the contention windows of all transmitters based on a novel,
    universally observable contention signal.

    \vspace{-2mm}
    \item We evaluate \sys using both simulations and commercial Wi-Fi APs,
    demonstrating that it significantly reduces Wi-Fi packet transmission latency and stabilizes throughput,
    ultimately reducing the video stall rate by over 90\%.
    % The code and data will be released after paper acceptance.
\end{itemize}

\vspace{-1mm}
\noindent\textbf{Ethical claim.} 
% All online data collected in this work are obtained with explicit permission from the users. This work does not raise any ethical concerns and conforms to the IRB policies of the authors' institutions.
All user data collected in this work are
obtained with explicit permission from the users and are
anonymized to protect their privacy. This work does not raise
any ethical concerns and conforms to the IRB policies of the
authors’ institutions.

%% file: sections/20_1_primer.tex
\section{Background }
To understand the latency challenges facing NGRTC applications,
this section provides essential background.
We first describe the architecture of these systems and their strict performance requirements.
We then examine the details of Wi-Fi's contention-based channel access,
which is the root of the performance bottlenecks we address.

\begin{figure}[t]
\setlength\abovedisplayskip{1pt}
\setlength\belowdisplayskip{1pt}
    \centering
    \includegraphics[width=0.9\linewidth]{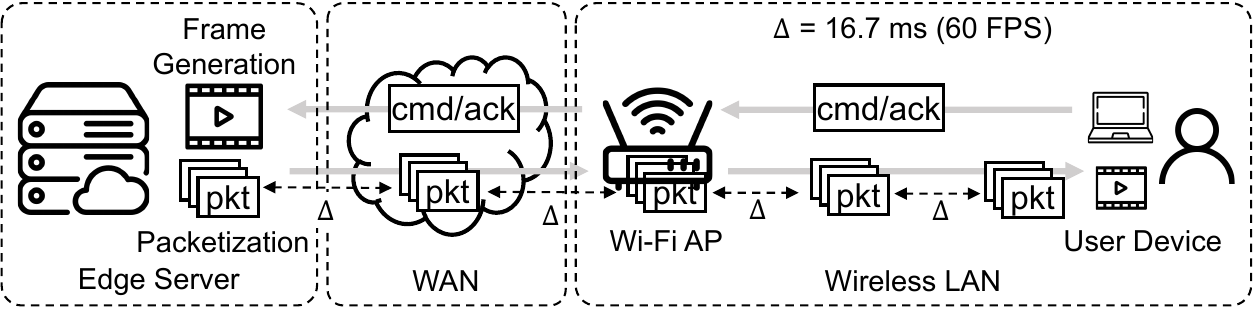}
    \caption{The system architecture of next-generation real-time streaming over wireless LAN.}
    \label{fig:rtc-arch}
    \vspace{-2mm}
\end{figure}

\subsection{Next-Generation RTC}
\label{sec:background-rt-streaming}

% Real-time streaming (RTS) over Wireless LAN (WLAN) has emerged as 
% the predominant choice for mobile gaming connectivity. 
% Based on our extensive analysis as a major commercial mobile gaming service provider, 
% we observe that wireless networks serve as the primary connection medium for 83.5\% of user sessions, 
% with Wi-Fi accounting for 56.7\% of all last-mile access connections. 
% This prevalence of WLAN usage extends beyond just connection statistics - our 
% data shows that users spend over 80\% of their total gaming time on Wi-Fi networks, 
% highlighting the critical role of WLAN performance in the user experience.
% \honghao{more than 80\% user time is relay on wifi} 
% \yx{Check?.}

\nosection{System Architecture.}
{
The inherent contradiction between computing power demands and terminal portability has
given rise to the core technical paradigm of computation and interaction
decoupling — computationally intensive tasks are processed in the cloud,
while terminals focus on low-latency local interaction and content presentation.
Typical applications such as cloud extended reality, AI smart glasses,
and cloud gaming are all built on this paradigm.
Compared with traditional Real-Time Communication (RTC) scenarios like live streaming and video conferencing,
these immersive interactive applications impose much more stringent performance requirements on network
transmission~\cite{shibo24-pudica, Latency1, latency2}: 
the end-to-end latency needs to be reduced from hundreds of milliseconds to tens of milliseconds,
and the bit-rate must be increased from several megabits per second to tens of megabits per second.
We define such dedicated network transmission demands for
immersive real-time interaction as Next-Generation Real-Time Communication (NGRTC). 
Based on our analysis at Tencent START cloud gaming service,
{\footnote{We infer access type from the client’s active network interface at session start and exclude sessions where access type is unknown.}}.
%Therefore, we focus on NGRTC over WLAN in this paper.
% This prevalence of WLAN usage extends beyond just connection statistics - our data shows that users spend over 80\% of their total gaming time on Wi-Fi networks, 
A typical NGRTC over WLAN system comprises four key components: 
the cloud server, WAN, Wi-Fi Access Point (AP), 
and user device, as illustrated in \figref{fig:rtc-arch}. 
The cloud server generates video frames at a fixed frame rate, and each frame is packetized into multiple packets for network transmission. 
For example, at 60 FPS, a new frame is generated and transmitted every 16.7 ms. 
These packets traverse the WAN to reach the Wi-Fi AP, 
which then delivers them wirelessly to the user's device. 
The system operates bidirectionally - upon receiving video frames, 
the user device sends acknowledgments (ACKs) along with 
interactive commands (\eg character movement or action triggers in mobile games) back to the cloud server. 
These user inputs then influence the generation of subsequent video frames.
}

\nosection{QoE Requirements on WLAN.}
The QoE of NGRTC critically depends on two key parameters:
video quality \cite{shibo24-pudica} and interaction smoothness \cite{2019-vrlatency, elbamby2018-vr-latency,
yuhan24-augur, shibo24-pudica, zili22-zhuge}.
For better video quality, these applications stream at much \textit{higher bitrates} (\eg over 30 Mbps for cloud gaming
\cite{shibo24-pudica} and over 200 Mbps for VR \cite{VR-200M}) compared to traditional RTC applications.
%NGRTC also requires a higher minimum bitrate than traditional RTC applications,
%with an additional 2--4 Mbps as the lower threshold \cite{geforce-now, start-tencent, xbox-cloud-gaming}.
For smooth interactions, they require a \textit{higher frame rate} 
(\ie 60 to 144 FPS) and demand \textit{consistently low video frame delivery latency}.
Specifically, \textit{tail latency} is particularly crucial: elevated tail latency--even at
99.99th percentile--can directly cause frequent video freezes and stalls \cite{shibo24-pudica,
yuhan24-augur}, significantly degrading user experience: a recent study~\cite{yuhan24-augur} has shown that
even a minor 0.5\% increase in stall rate leads
to a dramatic 33\% reduction in user retention time.
Thus, although Wi-Fi 6/7 offers theoretical rates of 9.6 Gbps and 46 Gbps,
far exceeding NGRTC's bitrate requirements,
NGRTC's sensitivity to tail latency imposes higher demands on
the real-time performance and stability of wireless network transmissions.

\subsection{WLAN Channel Access}\label{sec:background-wlan}
%WLAN operates as a decentralized system,
%where network devices compete for transmission opportunities 
%using a contention based medium access control protocol.
In this section, we introduce 
%the fundamental mechanisms of
Wi-Fi's contention-based channel access and packet transmission procedures.

\begin{figure}[t]
    \centering
    \includegraphics[width=0.95\linewidth]{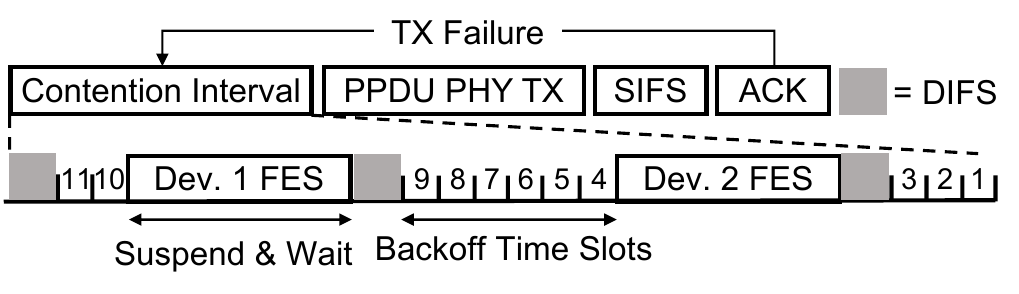}
    \caption{Wi-Fi Frame exchange sequence.}
    \label{fig:link-latency-decompose}
    \vspace{-2mm}
\end{figure}
\paragraph{Channel via CSMA/CA.}
Wi-Fi leverages \textit{carrier sense multiple access/collision avoidance} (CSMA/CA)
for channel access,
which requires a device to monitor channel activity before packet transmission.
Specifically, as illustrated in \figref{fig:link-latency-decompose},
a device must detect the channel as idle for $B$
\textit{backoff slots} before initiating transmission. The value of $B$ is randomly chosen from the range [0, CW] upon each transmission, where CW is the \textit{contention window} of the device.
If the device detects an ongoing transmission
during its countdown from $B$,
it suspends the countdown and resumes only after
detecting the channel as idle for a \textit{DCF~\footnote{DCF means 
distributed coordinate function in IEEE 802.11~\cite{ieee80211}.} interframe space} (DIFS) interval.
Upon successful completion of the $B$ slot countdown,
the device gains channel access for transmission.

\paragraph{Channel Contention Interval.}
We define the \textit{contention interval} as the period starting from DIFS until the successful completion
of $B$ backoff slots countdown.
During one device's contention interval,
other devices may gain channel access first,
as illustrated in \figref{fig:link-latency-decompose}.
Consequently, the duration of a contention interval is determined by two factors: the initial number of backoff slots ($B$)
and the number of channel access instances obtained by competing devices.

\paragraph{Wi-Fi Packet Transmission Procedure.}
After gaining channel access through CSMA/CA, the device encapsulates packets into \textit{PLCP Protocol Data Units} (PPDUs) with radio headers and proceeds with packet transmission during the
\textit{PHY TX} period shown in \figref{fig:link-latency-decompose}.
Upon successful PPDU reception,
the receiver must acknowledge by sending an ACK frame after a \textit{short inter-frame space} (SIFS) interval.
If a transmission fails (\ie no ACK or a NACK is received), the sender triggers a retransmission and re-gains channel access through CSMA/CA.
From the initial DIFS to the final ACK, it is defined as a \textit{frame exchange sequence} (FES) for a PPDU.

\paragraph{Wi-Fi MAC Throughput Analysis.}
Wi-Fi MAC throughput is determined by three key components: \textit{i)} PHY transmission rate, which dictates how quickly data packets can be transmitted over the air; \textit{ii)} Channel access overhead, including variable contention intervals and fixed intervals like DIFS, SIFS, and ACK; \textit{iii)} Transmission failure rate, failures are primarily caused by poor signal strength or signal collisions, where multiple devices attempt to transmit simultaneously.

\subsection{Predictability VS. Efficiency in CSMA}
It is important to distinguish the problem of tail latency from the well-studied issue of CSMA efficiency.
CSMA efficiency is typically defined as the ratio of airtime used for successful data
payload transmission to the total airtime consumed by a full frame exchange sequence,
which includes fixed overheads like the contention interval, DIFS, SIFS, and the ACK frame.
A long-standing challenge in Wi-Fi has been that as physical data rates increased,
the time to transmit a packet shrank, while these overheads remained constant,
causing a decline in overall channel efficiency.

This problem has been the subject of extensive research in the past two decades. 
Representative examples include fine-grained or frequency-domain contention to shrink time-domain overheads~\cite{Tan2010FICA,Sen2011NoTime}, 
explicit collision notification to curtail wasted airtime and hidden-terminal losses~\cite{Sen2010CSMACN}, 
and hybrid centralized--distributed coordination to exploit controller visibility while retaining CSMA agility~\cite{Shrivastava2009Centaur}. 
Earlier algorithmic tuning of contention parameters likewise sought high throughput and fairness under CSMA~\cite{Heusse2005IdleSense}. 
This problem is now largely
mitigated in modern Wi-Fi standards by highly effective solutions like frame aggregation 
(\textit{e.g.}, A-MPDU). 
By allowing multiple packets to be transmitted after a single contention event,
aggregation amortizes the overhead cost and significantly improves system throughput.
%While differing in mechanisms (\textit{e.g.}, AMPDU, OFDM subchannelization, frequency-domain backoff, collision notification, or controller-assisted scheduling), 
%these works share a common goal: improving \emph{efficiency} (throughput/airtime utilization) rather than controlling the distributional tail of per-packet access delays.

%
%
%Over the past two decades, most CSMA-oriented research has sought to raise \emph{efficiency} and
%throughput by reducing wasted airtime and extracting more payload per contention cycle.
%Representative examples include fine-grained or frequency-domain contention to shrink time-domain overheads~\cite{Tan2010FICA,Sen2011NoTime}, 
%explicit collision notification to curtail wasted airtime and hidden-terminal losses~\cite{Sen2010CSMACN}, 
%and hybrid centralized--distributed coordination to exploit controller visibility while retaining CSMA agility~\cite{Shrivastava2009Centaur}. 
%Earlier algorithmic tuning of contention parameters likewise sought high throughput and fairness under CSMA~\cite{Heusse2005IdleSense}. 
%While differing in mechanisms (e.g., OFDM subchannelization, frequency-domain backoff, collision notification, or controller-assisted scheduling), 
%these works share a common goal: improving \emph{efficiency} (throughput/airtime utilization) rather than controlling the distributional tail of per-packet access delays.
%

This paper, however, does not aim to solve the general CSMA efficiency problem.
Instead, we focus on a distinct but equally critical issue: the opportunistic
and severe inflation of the contention interval for specific packets.
While low efficiency is a systemic issue affecting average throughput,
the problem we address is transient and statistical. The long-term throughput
and average latency for a user can be perfectly acceptable,
yet the user's experience can be ruined by intermittent packet-delivery droughts where
the contention window for a single video frame inflates to an extreme value.
Efficiency is a problem of averages; 
contention-driven tail latency is a problem of outliers, and for NGRTC applications, these outliers are catastrophic.

%% file: sections/20_2_motivation.tex
\section{Measurement and Motivation}

In this section, we build the case for our proposed solution by first identifying and then
diagnosing the core performance problem for NGRTC applications in today's Wi-Fi networks.

\subsection{Large-Scale Online Measurement}\label{sec:online-measurement}
In this section, we present the first large-scale measurement study of Wi-Fi performance for NGRTC, 
revealing critical limitations in current Wi-Fi APs' ability to support these demanding workloads.

{While it is well known that CSMA/CA contention introduces variable per-packet delay, the application-level implications for modern high-bitrate interactive streaming---and the concrete failure mode that triggers stalls---have been unclear.
Our measurement contributes three pieces of evidence: (i) quantified stall-rate tails under Wi-Fi versus wired access, (ii) latency decomposition showing that the Wi-Fi last hop dominates even when WAN RTT is low and stable, and (iii) a near one-to-one correlation between 200~ms packet-delivery droughts and video stalls.
These findings motivate a link-layer contention-control mechanism that targets micro-level access fairness and bounds last-hop tail latency.}

\subsubsection{Testbed and Data Collection Scheme}
\nosection{Testbed.}
To understand how Wi-Fi last-hop performance impacts next-generation real-time streaming applications, 
we conducted an extensive measurement study through Tencent START cloud gaming platform. 
This platform delivers high-quality interactive gaming content, 
streaming 1080p to 4K video at 60-144 FPS with bitrates around 50 Mbps. 
Our measurement infrastructure consists of 200 commercial Wi-Fi access points distributed to volunteer users nationwide. 
{To reduce and stabilize server-side queuing delay (so that last-hop effects are more visible),
we deployed Pudica~\cite{shibo24-pudica} on the cloud-gaming servers.
Pudica enables near-zero queuing delay,
emerging as the state-of-the-art congestion control algorithm tailored for low-latency demands.}

%As shown in Figure~\ref{fig:acc_router_arch}, when the cloud gaming client establishes a connection with the server, it interacts with the router via the Router Interaction Module. This interaction triggers the Data Collection Module within the router, which retrieves relevant information, including Channel Status Information, MAC layer metrics, and PHY layer parameters, from the Wi-Fi driver. Subsequently, the Data Reporting Module within the router transmits the collected data to the server. The server then records and processes this data for further analysis. This architecture enables efficient data collection and reporting, facilitating seamless communication and monitoring between the client, router, and server components.

%\begin{figure}[t]
%    % \centering
%    \includegraphics[width=0.45\textwidth]{figure/acc_router_arch.pdf}
%    \caption{System Architecture for Data Collection from Wi-Fi Router. }
%    \label{fig:acc_router_arch}
%\end{figure}
\begin{figure}[t]
\centering
    \begin{minipage}[t]{0.49\linewidth}
    \centering
    \includegraphics[width=\textwidth]{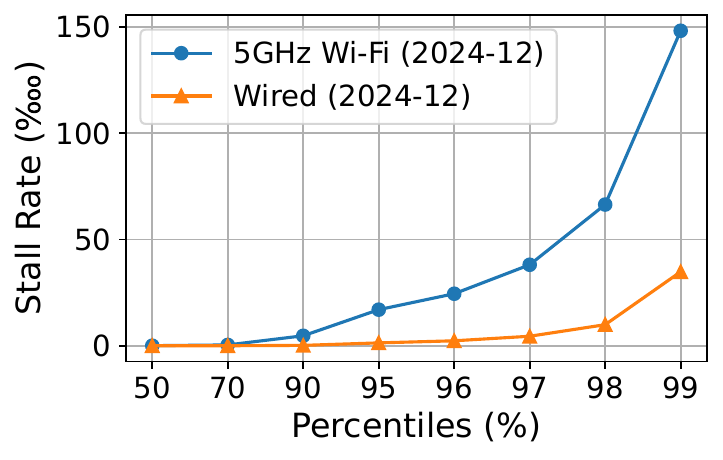}
    \caption{Stall rate percentiles in Dec. 2024. }
    \label{fig:stall-rate-1month}
    \end{minipage}
    \hfill
    \begin{minipage}[t]{0.49\linewidth}
    \centering
    \includegraphics[width=\textwidth]{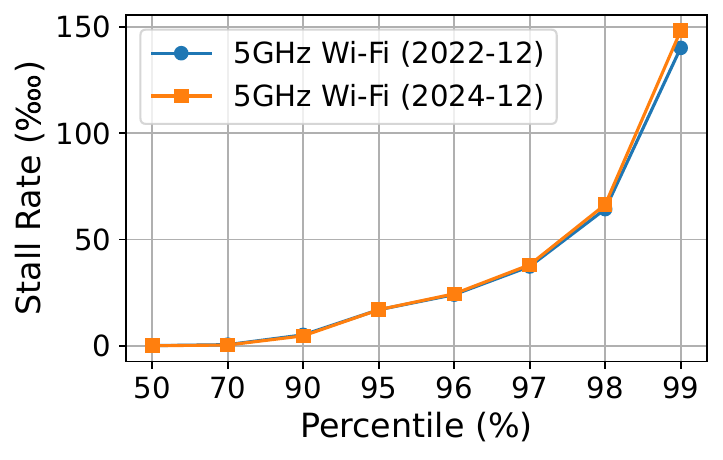}
    \caption{Stall rate for 5 GHz Wi-Fi in Dec. 2022 and 2024. }
    \label{fig:stall-rate-years}
    \end{minipage}
\end{figure}

\nosection{Data Collection.}
To collect comprehensive network performance data, we instrumented the WNIC driver on our Wi-Fi APs to report essential channel status, MAC layer metrics, and PHY layer parameters. 
The AP also records the successfully transmitted packets within each 200 ms interval,  
providing direct insight into wireless channel contention. 
Along with traditional metrics like RSSI, transmission delay, packet loss, and channel properties, 
the APs report these measurements every 200 ms.
This data collection scheme allows us to measure server-to-router RTT and distinguish between wired and wireless latency issues. Our server also collects transport layer statistics including frame-level RTT, packet loss, and jitter. 
{Concretely, the AP periodically measures a server\ensuremath{\leftrightarrow}AP RTT (every 200\,ms) over the control channel and reports it with the same granularity as MAC/PHY metrics. The server separately obtains per-frame end-to-end RTT from the cloud-gaming feedback path. We align the two by time (no clock synchronization is needed because both are RTTs).} Over one year, we gathered data from 336 million video frames---representing the first large-scale study of wireless last-hop performance for real-time streaming.

\subsubsection{Measurement Results}

\nosection{High Video Stall Rate.}
\figref{fig:stall-rate-1month} shows the video stall rate\footnote{We define a video stall as occurring when end-to-end frame delivery latency exceeds 200 ms. This metric is based on user QoE feedback and has been adopted by many previous studies in NGRTC \cite{shibo24-pudica, zili24-hairpin, yuhan24-augur}.} percentiles for cloud gaming users in December 2024 across different networks. {We report stall rate as stalls per 10{,}000 frames (\ensuremath{\times}10\ensuremath{^{-4}}), so values above 100 correspond to more than 1\% of frames stalling.} 5 GHz Wi-Fi exhibits significantly higher tail latency compared to wired networks, indicating the superior stability of wired connections.
{\figref{fig:stall-rate-years} compares the stall-rate percentiles for 5~GHz Wi-Fi sessions from two matched one-month snapshots (Dec.~2022 vs.~Dec.~2024) under the same stall definition.}
{The similarity indicates that, even as Wi-Fi hardware evolves, tail stalls driven by CSMA/CA contention remain a dominant factor in dense environments.}

\begin{figure*}[t]
\centering
    \begin{minipage}[t]{0.24\linewidth}
    \centering
    \includegraphics[width=\textwidth]{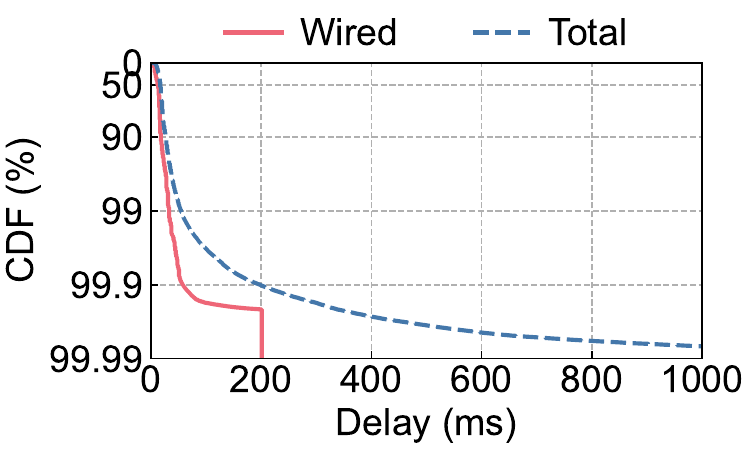}
    \caption{Distribution of video frame latency in cloud gaming.}
    \label{fig:ap-mirror-cdf}
    \end{minipage}
    \hfill
    \begin{minipage}[t]{0.24\linewidth}
    \centering
    \includegraphics[width=\textwidth]{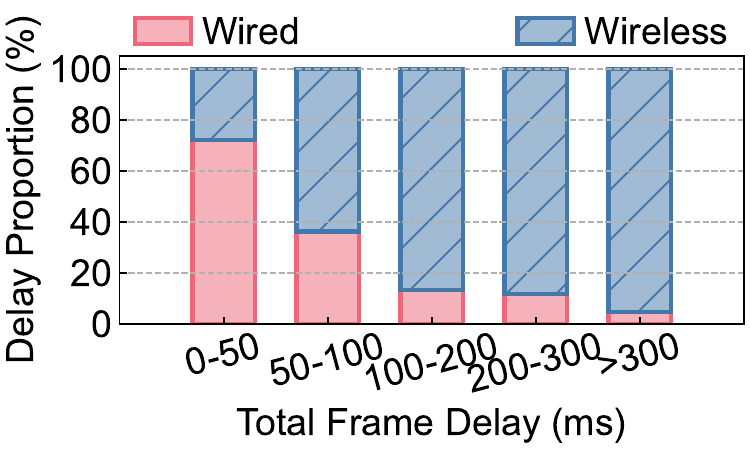}
    \caption{Cloud gaming video frame latency decomposition.}
    \label{fig:ap-mirror-decomp}
    \end{minipage}
    \hfill
    \begin{minipage}[t]{0.24\linewidth}
    \centering
    \includegraphics[width=\textwidth]{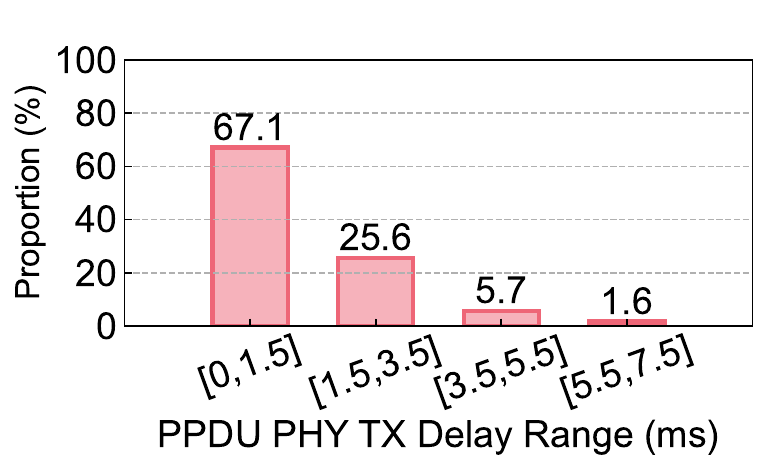}
    \caption{Distribution of Wi-Fi PHY transmission delay.}
    \label{fig:online-phy-tx}
    \end{minipage}
    \hfill
    \begin{minipage}[t]{0.24\linewidth}
    \centering
    \includegraphics[width=\textwidth]{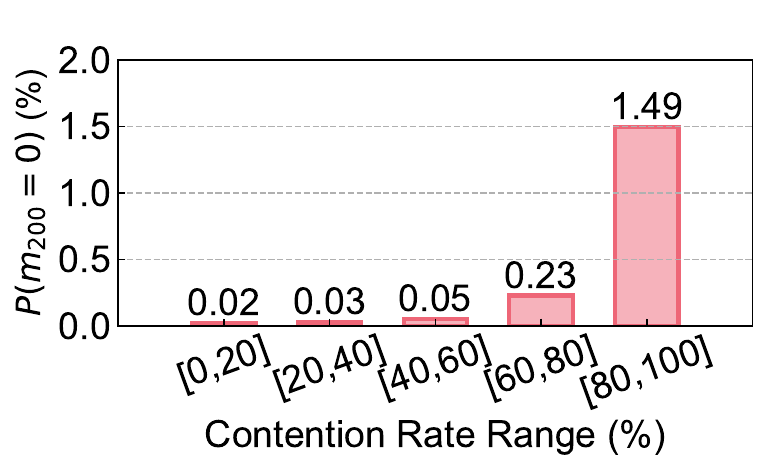}
    \caption{The distribution of transmission opportunity.}
    %under different channel contention rates.}
    \label{fig:online-contention-stall}
    \end{minipage}
    \vspace{-2mm}
\end{figure*}

\nosection{Wi-Fi Last-Mile Causes High Video Stall Rate.}
Our measurement campaign reveals the root cause of these video stalls: 
the Wi-Fi last hop acts as a critical performance bottleneck. 
By analyzing each video frame's end-to-end path, 
we find striking differences between wired and wireless segments. 
As shown in \figref{fig:ap-mirror-cdf}, 
the wired portion (server to AP) maintains consistently low latency, 
staying below 200 ms even at the 99.99th percentile. However, 
when including the wireless last hop (AP to user), 
the total latency can exceed 1000 ms.
To precisely quantify this impact, we decomposed each frame's delivery time into wired and wireless components. 
\figref{fig:ap-mirror-decomp} reveals that the wireless segment contributes disproportionately to total latency, 
with its share growing dramatically as delivery times increase. 
This finding is particularly concerning because it shows that even with state-of-the-art WAN congestion control, 
the wireless last hop remains the primary obstacle to reliable real-time streaming.

%\nosection{Wi-Fi Last-Mile Dominates Frame Delivery Latency.}
%Our measurement reveals that the Wi-Fi last hop serves 
%as the primary bottleneck for video frame delivery, 
%significantly inflating tail latency. 
%Figure~\ref{fig:ap-mirror-cdf} compares the wired latency (server to AP) against the total latency (server to user device) for all video frames. 
%While the wired segment maintains consistent performance with latency below 200ms even at the 99.99th percentile, 
%the total latency can exceed 1000ms, indicating substantial delays introduced in the wireless last hop. 
%To quantify this impact, we further decompose the latency into wired and wireless components for each frame. As shown in Figure~\ref{fig:ap-mirror-decomp}, the wireless segment's contribution to total latency grows disproportionately as frame delivery time increases, ultimately becoming the dominant factor in degraded performance. This finding highlights that despite optimized WAN transport with state-of-the-art congestion control, the wireless last hop remains a critical bottleneck for real-time streaming applications.
%
\begin{table}[t]
    \centering
    \small % 调整字体大小为小号
    \setlength{\tabcolsep}{5pt} % 调整列间距
    \renewcommand{\arraystretch}{1} % 调整行间距
    \begin{tabular}{cc|cc}
        \Xhline{1.5\arrayrulewidth}
        \textbf{Range} & \textbf{Probability (\%)} & \textbf{Range} & \textbf{Probability (\%)} \\
        \Xhline{1.5\arrayrulewidth}
        0 & 86.19 & 5 & 0.78 \\
        1 & 0.29 & [6,10) & 2.55 \\
        2 & 0.39 & [10,20) & 2.86 \\
        3 & 0.36 & [20,50) & 2.46 \\
        4 & 0.29 & (50,$\infty$) & 3.82 \\
        \Xhline{1.5\arrayrulewidth}
    \end{tabular}
    \vspace{2mm}
    \caption{The distribution of the number of packets transmitted by the Wi-Fi router within 200 milliseconds, when downstream long-tail latency occurs (with the absence of wired network issues confirmed).}
    \label{tab:m200-prob}
\end{table}

\nosection{Packet Delivery Droughts: Root Cause of Frame Stalls.}
To identify the root cause of Wi-Fi last-hop delays, we analyzed the correlation 
between stalled video frames and successful packet transmissions. 
For each frame with a high end-to-end delay (server to client), 
we examined the number of successfully transmitted packets within each 200 ms window of the frame's transmission. 
To isolate Wi-Fi-induced stalls, we focused on frames where server-to-client latency exceeded 200 ms 
while server-to-router RTT remained below 50 ms---effectively filtering out stalls caused by wired network issues.

Table~\ref{tab:m200-prob} reveals a striking pattern: in 86.19\% of these stalled frames, 
the router failed to successfully transmit even a single packet during at least one 200 ms interval, 
despite potentially having transmission opportunities. 
This near one-to-one correspondence between packet delivery droughts and 
frame stalls suggests a fundamental issue in Wi-Fi's channel access mechanism, 
where either transmission opportunities are not obtained or packets fail to be delivered even when opportunities are granted. In contrast, we calculate the PHY transmission delay of PPDUs and present its distribution in \figref{fig:online-phy-tx}. Once PPDUs are granted transmission opportunities, the actual transmission completes quickly, with 92.7\% finishing within 3.5 ms and a maximum delay of 7.5 ms.

To further understand why AP fails to deliver packets, 
we investigated the relationship between packet delivery and channel contention. 
We define the channel contention rate as the proportion of airtime 
occupied by other transmitters within each 200 ms interval (longer airtime by others indicating higher contention). 
\figref{fig:online-contention-stall} shows that the probability of 
zero packet deliveries rises dramatically with increased 
channel contention---when contention exceeds 80\%, 
the probability of a complete delivery drought is 74.5 times higher than under 20\% contention.

We further validated this relationship through an 8-week field study where, with user consent, 
we monitored the number of nearby Wi-Fi APs as a proxy for potential channel contention. 
Table~\ref{tab:online-apnum-stall}  demonstrates that video stall rates, 
particularly at the tail, increase systematically with the number of surrounding APs. 
These findings establish that frame stalls primarily occur 
when routers experience packet delivery droughts during periods of intensive channel contention.
{Importantly, a single 200\,ms delivery drought already crosses the stall threshold used by the application, so mitigating micro-level droughts is directly reflected in lower stall rate and better user QoE.}

\begin{table}[t]
    \centering
    \resizebox{0.8\linewidth}{!}{
    \begin{tabular}{cccc}
    \Xhline{2\arrayrulewidth}
    \textbf{AP Num.} & \textbf{Session Num.} & \textbf{Stall Rate (\%)} \\
    \Xhline{2\arrayrulewidth}
    2 & 52349 & 0.08 \\
    4 & 25624 & 0.17 \\
    6 & 14414 & 0.42 \\
    $\ge 8$ & 7976 & 1.34\\
    \Xhline{2\arrayrulewidth}
    \end{tabular}
    }
    \vspace{3mm}
    \caption{Relations between video stall rate of Wi-Fi sessions and Wi-Fi AP numbers in the environment from our online cloud gaming platform for 8 weeks.}
    \label{tab:online-apnum-stall}
% \vspace{-8mm}
\end{table}

\subsection{Mechanism Behind Packet Delivery Droughts}\label{sec:wifi-limit}
{Our online measurements establish that delivery droughts concentrate under high contention, but they do not expose the packet-level dynamics that create 100--200\,ms gaps. We therefore complement them with ns-3 simulations and controlled experiments with commercial Wi-Fi APs. Across both settings, we find that (i) collisions increase retransmissions, and (ii) each retransmission triggers binary exponential backoff whose countdown is repeatedly frozen under a busy channel, stretching the \emph{effective} contention interval from sub-millisecond to hundreds of milliseconds. We report the full methodology and supporting figures in \secref{append:drought-mech}. These dynamics point to a deeper limitation: \emph{802.11’s contention control is collision-driven and purely reactive}, which we summarize next.}

\subsubsection{Root Cause: Collision-Driven Reactive Contention}
The fundamental issue lies in 802.11's reactive approach to contention control. As detailed in \secref{append:drought-mech}, the long gaps are dominated by collision-driven retransmissions and prolonged countdown freezes, rather than PHY transmission time.
Current CSMA/CA mechanism has two critical limitations that lead to extended packet delivery times.
First, the protocol always initializes transmission with a small contention window ($CW_{min}$), 
regardless of network contention levels. 
In dense networks with high contention, 
this approach inevitably leads to frequent collisions---multiple devices are 
likely to select similar small backoff values. 
A more effective approach would be to proactively adjust the 
initial window size based on observed network contention, 
starting with larger windows when contention is high.

Second, the protocol creates unfair channel access after collisions. 
When a device experiences a collision, it doubles its contention window, 
while devices without recent collisions maintain small windows. 
This creates a problematic asymmetry: devices with larger windows must count down through more slots, 
making them more likely to be interrupted by transmissions from devices with smaller windows. 
Each interruption forces the device to pause its countdown until the channel becomes idle again. 
In dense networks, these interruptions can extend a simple backoff period from milliseconds to hundreds of milliseconds.

This reactive, device-by-device approach means the system never 
achieves a coordinated response to network contention. 
Instead, devices independently adjust their windows based only on their own collision experiences, 
leading to persistent unfairness and inefficient channel utilization. 
A better approach would be to maintain similar window sizes across all devices 
based on overall network contention levels, 
ensuring fair channel access while proactively preventing collisions.

%% file: sections/40-design.tex
\section{\sys Design}
To meet the latency and throughput stability requirements of NGRTC in Wi-Fi networks, we consider the following two aspects as the most critical: (1) maintaining a low collision probability; (2) ensuring that the CW of different co-channel Wi-Fi devices remains as consistent as possible. Based on these two principles, we design \sys, which leverages MAR and employs the HIMD approach to adaptively adjust CW, achieving a balance between high throughput and low collision probability, without relying on priority queues in Wi-Fi networks. We first present the design goals of \sys.

% \sys adjusts each transmitter's contention window value in a fully distributed way to ensure micro-level bandwidth fairness for all transmitters. 

% with an easy-to-get input signal (\secref{sec:signal}). \sys is designed to be simple and robust, its control logic consists of two components: \textit{i)} A \textit{stable-state control policy} (\secref{sec:stable-control}) that maintains the contention windows of all transmitters adaptive to the channel contention level and ensures rapid convergence; \textit{ii)} A \textit{transient-state control policy} (\secref{sec:transient-control}) designed to handle occasional transmission failures due to collisions. We first present the design goals and challenges behind \sys.

\subsection{Design Goals}
A practical and effective contention window adjustment mechanism should achieve four key goals:

\paragraph{High Transmission Efficiency.}
The system must maximize channel utilization by maintaining an optimal collision rate. 
Since collision recovery (3-5ms) costs significantly more than contention slot time (9$\mu$s), 
we need to balance between avoiding excessive collisions from small contention windows 
and preventing unnecessary idle periods from large windows.

\paragraph{Fair Channel Access.}
All devices should reach a consensus on network contention levels and adjust their windows accordingly. 
This differs from current IEEE mechanisms where devices react individually 
to their own transmission outcomes, leading to unfair access patterns. 
Instead, devices should collectively adapt their contention windows 
based on shared network conditions, ensuring balanced transmission opportunities across the network.

%All devices should have equal transmission opportunities regardless of their transmission history. 
%This differs from current IEEE mechanisms where successful transmissions lead to 
%small windows (and thus more access) while collisions trigger exponential backoff. 
%The system should enable synchronized window adjustments 
%across all devices based on network conditions.

\paragraph{Fast Convergence.}
The system should rapidly adapt to network changes 
while maintaining stable operation. 
When network conditions shift (\eg, traffic flows joining or leaving), 
all devices should quickly converge to appropriate window sizes, 
maintaining both efficiency and fairness in their channel access patterns 
without oscillating between states.

\nosection{Minimal Assumptions.}
The system should be designed without relying on assumptions about user traffic patterns, the number of competing flows, or PPDU PHY transmission duration, as real-world networks are inherently complex. They exhibit unpredictable user traffic, highly dynamic competing transmitters, and varying PHY transmission rates. This contrasts with existing studies on contention window control algorithms \cite{heusse05-idlesense, yang06-dda, tian2005-dob, qin12-dcwa, qiang03-slowdec} beyond the IEEE 802.11 standard, which are based on these assumptions.

% \yuhan{Zero assumption?}

These goals are particularly challenging because Wi-Fi 
operates as a fully distributed system where devices must make decisions without explicit coordination.

\subsection{Search of a Universal Contention Signal} \label{sec:signal}
%\subsubsection{Requirements for Network Contention Signal}

\paragraph{Requirement.}
To enable coordinated contention window adjustment across distributed Wi-Fi devices, 
we need a reliable signal that indicates network contention levels. 
This signal must satisfy three key requirements: 
it should be universally observable by all devices, 
accurately reflect current network competition, 
and remain stable enough to facilitate consensus.
Several candidate signals face fundamental limitations:
\textit{i)} Collision-based signals only provide local feedback to involved devices;
\textit{ii)} Detecting competing flows requires packet-level decoding at the MAC layer, which is both complex to implement and potentially misleading—flows 
operate at longer time scales and may be temporarily inactive, 
making them poor indicators of instantaneous network contention;
\textit{iii)} Airtime utilization rate (the fraction of airtime occupied by transmissions) can be deceptive, since high utilization rate may simply caused by large PPDUs from few devices rather than actual competition for channel access.

\subsubsection{Proposed Signal: MAR}\label{sec:mar-def}

%We propose that the signal utilized for contention window (CW) adjustment should be directly correlated with the degree of competition for transmission opportunities. Based on this premise, we advocate for the adoption of MAR as the signal for CW adaptation. 

\begin{figure}
    \centering
    \includegraphics[width=\linewidth]{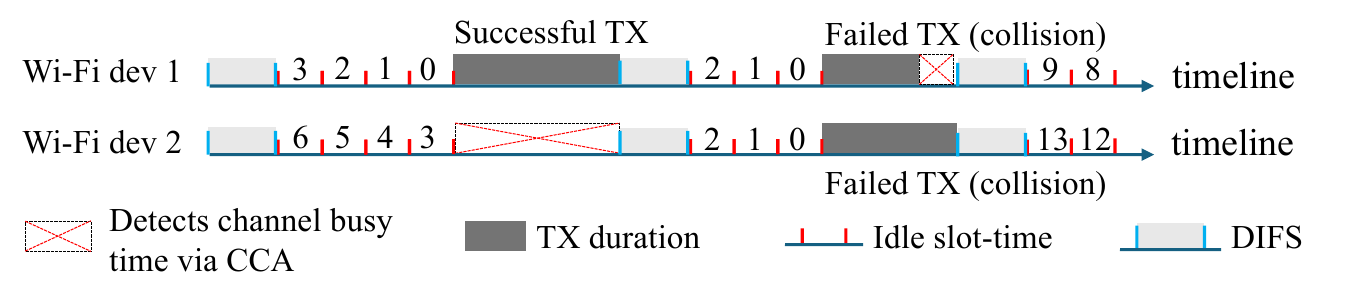}
    \vspace{-4mm}
    \caption{Illustration of MAR. There are 9 idle slot times (in red) and 2 TX durations, the MAR that both device 1 and device 2 detect via CCA is $2/(9+2)$.}
    \label{fig:tx-opportunity}
\end{figure}

\paragraph{Definition of MAR.}
We define the \textit{microscopic access rate} (MAR) as the ratio of 
transmission opportunities to total available slots in the channel. 
As shown in \figref{fig:tx-opportunity}, 
each device monitors both idle slots during its backoff countdown and transmission events in the channel. 
A transmission event occurs either when the device itself gains channel access 
or when it detects other devices' transmissions through CCA. 
Mathematically, MAR is defined as:
\begin{equation}\label{MAR_def}
\setlength\abovedisplayskip{1pt}
\setlength\belowdisplayskip{1pt}
    MAR = \tfrac{N_{tx}}{N_{tx} + N_{idle}}
\end{equation} 
where $N_{tx}$ is the number of transmission events 
and $N_{idle}$ is the number of idle slots during backoff countdown. 
In the example shown in \figref{fig:tx-opportunity}, 
there are 2 transmission events and 9 idle slots, resulting in a $MAR$ of 2/11. 

% Importantly, this metric is consistently observable by all devices sharing the same channel, as they all freeze their countdown during others' transmissions and count the same idle slots.

\paragraph{MAR: Properties and Advantages.}
MAR offers three key advantages as a contention signal:

\vspace{0.03cm}
\noindent
\textit{Universal Observability.}
{For devices that can carrier-sense each other (i.e., within the same carrier-sense domain), MAR is consistently observable: when any device transmits, others detect it via CCA and freeze backoff, leading to a shared sequence of transmission events and idle slots. Hidden terminals and partial visibility can violate this assumption; we discuss mitigation via RTS/CTS and empirically validate robustness in \secref{append:hidden-terminal}.}

\vspace{0.03cm}
\noindent
\textit{Direct Competition Indicator.}
MAR directly reflects the intensity of channel competition by measuring the ratio of transmission attempts to available slots. Unlike network utilization or flow counts, MAR captures the actual contention for transmission opportunities, allowing devices to accurately gauge network competition levels.

\vspace{0.03cm}
\noindent
\textit{Predictable Collision Control.}
When devices maintain MAR at a target threshold through contention window adjustment, collision probability remains stable regardless of the number of competing devices (See \secref{sec:Collision Probability upper bound} for proof). This property enables systematic congestion management without requiring knowledge of network size or traffic patterns.

\subsection{MAR-Driven Contention Window Control}\label{sec:stable-control}
\paragraph{Problem Statement}
The core challenge in MAR-driven contention control is to 
dynamically adjust each transmitter's contention window (CW) to achieve three key objectives. 
\textit{i)} the system must maintain the observed MAR close to a 
target value $MAR_{tar}$ to ensure efficient channel utilization;
\textit{ii)} all competing transmitters must converge to similar CW values to 
guarantee fair channel access---significant differences in CW values would give some 
transmitters unfair advantages in channel competition. 
\textit{iii)} the system needs to rapidly adapt CW values in response to network changes, 
converging quickly to optimal settings without oscillation.

These objectives present inherent tensions. 
Aggressive CW adjustments achieves faster convergence but risks creating temporary unfairness or oscillations. 
Conservative adjustments provide more stability but may react too slowly to network changes. 
Additionally, transmitters must achieve these objectives through independent decisions 
without explicit coordination, as the distributed nature of Wi-Fi networks precludes direct communication between devices.

\subsubsection{HIMD-based Contention Window Control}\label{sec:stable-control}
Drawing inspiration from traditional TCP congestion control, 
we design a \textit{hybrid increase multiplicative decrease} (HIMD) policy for CW adjustment. 
{Note that the ``increase/decrease'' directions are inverted compared to transport-layer congestion windows: in Wi-Fi, a larger contention window reduces a transmitter's attempt probability, so ``increasing CW'' makes the transmitter \emph{less} aggressive.}
Traditional AIMD (additive increase multiplicative decrease) 
has been proven to achieve fair bandwidth sharing in congestion control. 
We extend it with a hybrid increase phase 
that combines both additive and multiplicative components---the additive 
component ensures steady fairness convergence, 
while the multiplicative component provides rapid response to severe congestion. 
This hybrid approach offers better adaptivity than pure AIMD while maintaining its fairness properties. 
Like AIMD, our HIMD policy increases CW when MAR exceeds $MAR_{tar}$ to reduce channel contention, 
and decreases CW when MAR is below $MAR_{tar}$ to encourage more transmission attempts. {Here, $MAR_{tar}$ is the target microscopic access rate that we regulate to in steady state (default 0.1), and $MAR_{max}$ is an empirical upper bound of MAR under saturated contention (default 0.35), used to normalize/clip the control signal and avoid over-reacting when the channel is nearly fully occupied by transmissions and fixed MAC overheads.}

%To maintain high transmission efficiency and avoid collision, \sys observes MAR periodically with an observation interval $N_{obs} = N_{tx} + N_{idle}$. We calculate $N_{obs} = 300$ to be an appropriate value and leave the detailed analysis in \secref{appendix:observation-interval}. Based on the observed MAR, \sys adapts the CW value of all transmitters to the channel contention level, converging the system to a stable state. Specifically, \sys adopts an \textit{Additive and Multiplicative Increase, Multiplicative Decrease (AMI-AD)} approach.

\nosection{Hybrid Increase.}
When the observed $MAR$ exceeds the target $MAR_{tar}$, it indicates excessive contention, leading to more collisions and longer contention intervals. To alleviate this, \sys increases the contention window $CW$ to yield more transmission opportunities and reduce contention:
{\begin{align} 
\setlength\abovedisplayskip{1pt}
\setlength\belowdisplayskip{1pt}
    CW = CW &+ M_{inc}(\min\{MAR, MAR_{max}\} - MAR_{tar}) + A_{inc} \nonumber \\
    &+ CW \cdot \max\{0, MAR - MAR_{max}\} \label{eqn:ami}
\end{align}}
\eqnref{eqn:ami} involves two additive and one multiplicative terms: \textit{i)} Additive term $M_{inc}(\min\{MAR, MAR_{max}\} - MAR_{tar})$ ensures a faster increase in CW when the observed MAR significantly exceeds the target and a slower increase when it is close to the target. We use $M_{inc} = (CW_{max} - CW_{min}) / 2$ by default; \textit{ii)} Additive term $A_{inc}$ guarantees a minimum increase, promoting fairness among all transmitters' $CW$ values; \textit{iii)} Multiplicative term $CW \cdot \max \{0, MAR - MAR_{max} \}$ is applied to handle extreme contention scenarios. When the observed MAR exceeds $MAR_{max}$, the channel is considered highly congested and unstable. 
In response, CW is increased multiplicatively to rapidly reduce contention.
{Overall, \eqnref{eqn:ami} behaves as a stable proportional controller on the MAR error within a safe range, while providing (i) a fairness floor via $A_{inc}$ and (ii) an emergency brake via the multiplicative term when the system enters a highly congested regime ($MAR>MAR_{max}$).}
The default value of $MAR_{max}$ is set to 35\%, as our simulation experiments revealed that under the IEEE standard, 
the MAR tends to rise to approximately 35\% with an increasing number of competing Wi-Fi flows. %\yuhan{any proof?}

\nosection{Multiplicative Decrease.}
When the observed $MAR$ is below the target $MAR_{tar}$, 
it indicates insufficient traffic load, wasting transmission chances and overall bandwidth. 
To effectively utilize the airtime resource, \sys should rapidly contend 
for more transmission chances by decreasing the contention window $CW$ multiplicatively: $CW = \beta \cdot CW, \beta < 1$.
Our choice on $\beta$ value involves two factors:
On the one hand, to quickly converge without oscillation, we aim for the observed $MAR$ to increase by $(MAR_{tar} - MAR) / 2$ per step. {In the converged state, $MAR$ is (approximately) inversely proportional to the converged $CW$: with attempt probability $\tau\approx \tfrac{2}{CW+1}$ per transmission chance, $MAR = 1-(1-\tau)^N\approx N\tau\approx \tfrac{2N}{CW+1}$ for $\tau\ll 1$.} Therefore, we use
\begin{equation}
\setlength\abovedisplayskip{1pt}
\setlength\belowdisplayskip{1pt}
    \beta_1 = \tfrac{MAR}{MAR_{tar} - \tfrac{MAR_{tar} - MAR}{2}} = \tfrac{2MAR}{MAR_{tar} + MAR}
\end{equation}
On the other hand, to accelerate fair convergence, the greater the CW value is, the larger the reduction magnitude should be, therefore, we use
\begin{equation}
\setlength\abovedisplayskip{1pt}
\setlength\belowdisplayskip{1pt}
    \beta_2 = M_{dec} - \tfrac{(1 - M_{dec})(CW - CW_{min})}{CW_{max} - CW_{min}}\label{beta_2}
\end{equation}
where $M_{dec}$ is a minimum decrease factor, with a default value of 0.95. The second term on the right-hand side of \eqnref{beta_2}  ensures that Wi-Fi devices with larger $CW$ values experience a greater reduction, thereby speeding up the convergence process. 
Finally, combining the two considerations, we update the contention window as:
\begin{equation}
\setlength\abovedisplayskip{1pt}
\setlength\belowdisplayskip{1pt}
    CW = \min(\beta_1, \beta_2)\cdot CW
\end{equation}
{Because $MAR$ is a channel-wide consensus signal (all transmitters on the same channel observe the same busy/idle pattern), all nodes react to a common feedback loop. $\beta_1$ drives the system toward the fixed point where $MAR\approx MAR_{tar}$, while $\beta_2$ contracts CW disparities faster by applying larger reductions to larger $CW$ values. Taking $\min(\beta_1,\beta_2)$ avoids overshooting and reduces oscillation. Together with hard bounds $[CW_{min},CW_{max}]$, this yields rapid convergence without persistent unfairness.}

\nosection{Target MAR.}
{The target microscopic access rate $MAR_{tar}$ is a critical parameter in our HIMD control policy.} {Using a standard CSMA/CA throughput model, the throughput-optimal $MAR$ is approximately $MAR_{opt}=\tfrac{1}{\sqrt{\eta}+1}$, where $\eta\!=\!T_c/T_s$ is the collision duration (in slots) relative to an idle backoff slot.} {In modern Wi-Fi, $\eta$ is typically large (collisions last tens to hundreds of slots), which places $MAR_{opt}$ in a narrow ``safe'' band around 0.1.} {Accordingly, we set $MAR_{tar}=0.1$ by default, and \secref{sec:microbench-util-rate} shows that \sys remains robust when $MAR_{tar}$ varies within this band.}

%$\nosection{Target MAR.}
%$The target microscopic access rate $MAR_{tar}$ plays an essential
%$role in \sys's stable state control policy. By default, we set $MAR_{tar} = 0.1$, we argue that such a value is optimal. More importantly, \sys's performance remains largely stable when $MAR_{tar}$ deviates from this optimal value. We leave the detailed analysis on $MAR_{tar}$ in \secref{appendix:design-theory} and further evaluate the impact of $MAR_{tar}$ in \secref{sec:microbench-util-rate}.
%$
%\yuhan{How to explain 0.05?}
\paragraph{Fast Recovery Policy for Collisions.}
While our HIMD policy ensures stable convergence, random collisions can still occur when multiple transmitters select the same backoff value. To minimize the delay impact of these collisions, we implement a special handling for retransmissions. Upon a transmission failure, instead of the standard IEEE 802.11 approach of doubling CW, we set:
\begin{equation}
\setlength\abovedisplayskip{1pt}
\setlength\belowdisplayskip{1pt}
CW_{fail} = CW + A_{fail} \
CW = CW_{fail} / 2
\end{equation}
This temporary CW reduction accelerates retransmission of collided packets while $A_{fail}$ serves as a compensation term. After successful retransmission, we restore CW to $CW_{fail}$ before resuming normal HIMD control. To prevent excessive contention, this halving is applied only to the first retransmission attempt.

%% file: sections/50-implmentation.tex
\section{Implementation}\label{sec:impl}
We implement \sys on Tenda AX12 Pro Wi-Fi APs, accessing the Wi-Fi driver layer to monitor channel activity. 
Our implementation primarily leverages three hardware counters from the CCA mechanism:
\textit{TX\_time}: duration of AP's active data transmission;
\textit{BUSY\_time}: duration when channel is busy with other transmissions;
\textit{IDLE\_slot\_time}: count of idle channel slots.
We poll these microsecond-precision counters every 1 ms and calculate the observed MAR by tracking changes in counter values. This provides accurate measurement of $N_{tx}$ and $N_{idle}$ as defined in \secref{sec:mar-def}.
For CW control, we implement the complete HIMD algorithm with: 
observation interval of 300 slots when calculating MAR (justified in \secref{appendix:observation-interval})
and standard BE queue parameters ($CW_{min}=15$, $CW_{max}=1023$).
The implementation consists of approximately 500 lines of C code, 
focusing on counter monitoring and CW adjustment logic.

%% file: sections/60-evaluation.tex
\section{Evaluation}
We evaluate \sys's performance through extensive experiments on commercial Wi-Fi APs and ns3 simulations.
The assessment covers diverse network conditions, from saturated links to realistic traffic,
culminating in real-world tests with cloud gaming applications.
We benchmark \sys against the standard IEEE 802.11 contention control
and other relevant algorithms to validate its effectiveness.

\begin{figure*}[t]
\centering
    \begin{subfigure}[t]{0.245\linewidth}
	\centering
	\includegraphics[width=\textwidth]{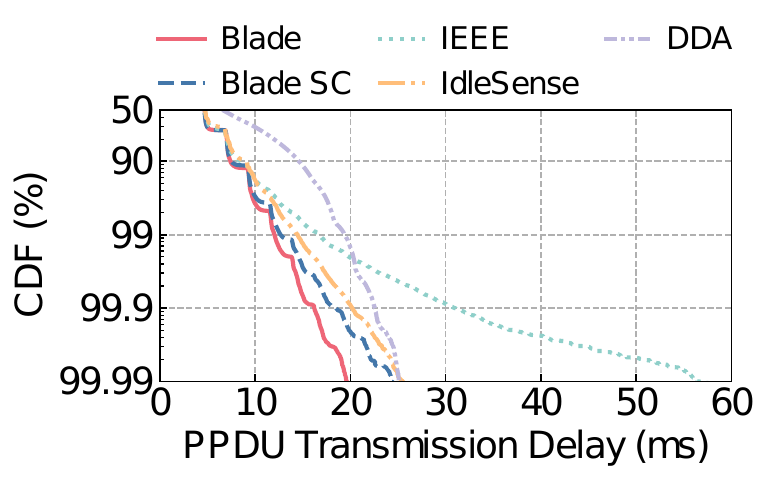}
	\caption{$N = 2$.}
	\label{fig:sim-wild-hol-2}
    \end{subfigure}
    \hfill
    \begin{subfigure}[t]{0.245\linewidth}
	\centering
	\includegraphics[width=\textwidth]{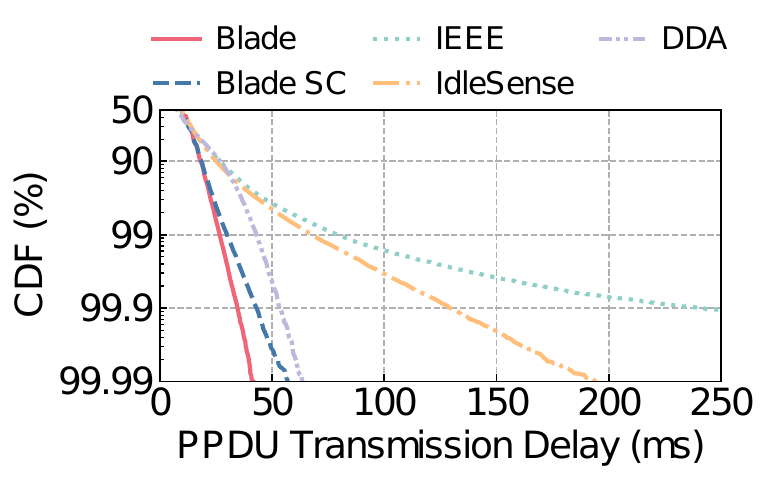}
	\caption{$N = 4$.}
	\label{fig:sim-wild-hol-4}
    \end{subfigure}
    \hfill
    \begin{subfigure}[t]{0.245\linewidth}
	\centering
	\includegraphics[width=\textwidth]{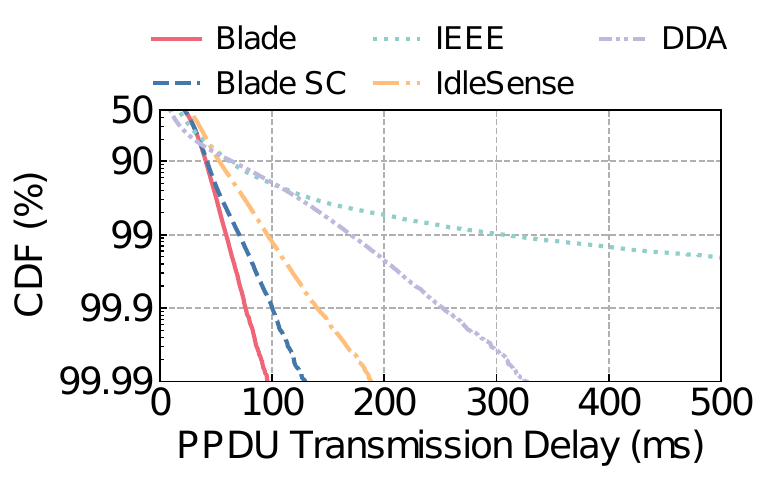}
	\caption{$N = 8$.}
	\label{fig:sim-wild-hol-8}
    \end{subfigure}
    \hfill
    \begin{subfigure}[t]{0.245\linewidth}
	\centering
	\includegraphics[width=\textwidth]{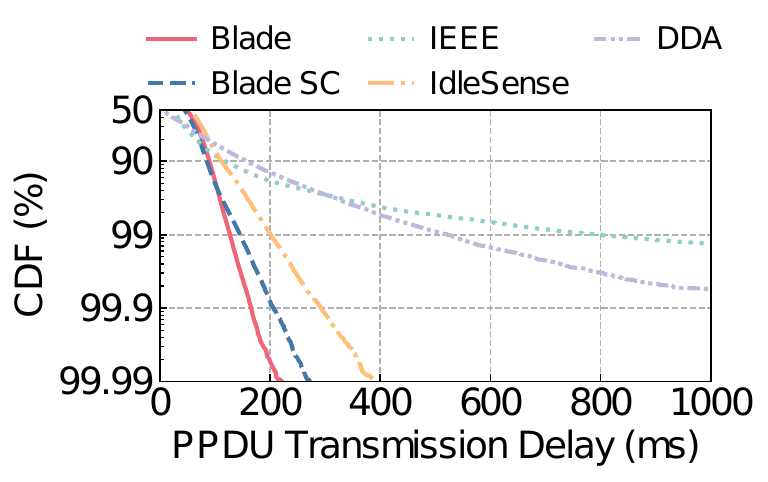}
	\caption{$N = 16$.}
	\label{fig:sim-wild-hol-16}
    \end{subfigure}
    \caption{PPDU transmission delay distribution under $N$ competing flows.}
    \label{fig:sim-wild-hol}
    \vspace{-2mm}
\end{figure*}

\begin{figure*}[t]
\centering
    \begin{subfigure}[t]{0.245\linewidth}
	\centering
	\includegraphics[width=\textwidth]{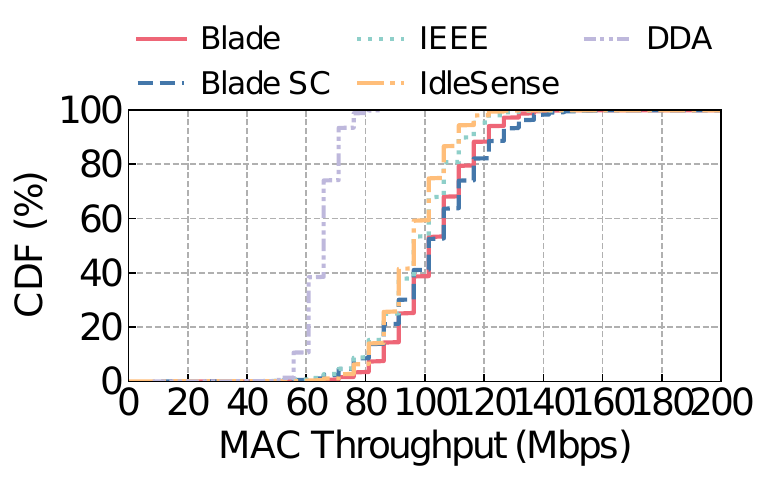}
	\caption{$N = 2$.}
	\label{fig:sim-wild-mac-thp-2}
    \end{subfigure}
    \hfill
    \begin{subfigure}[t]{0.245\linewidth}
	\centering
	\includegraphics[width=\textwidth]{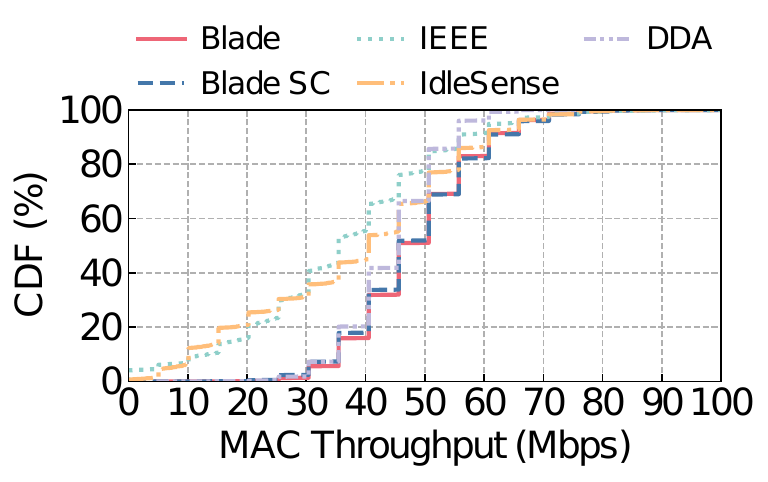}
	\caption{$N = 4$.}
	\label{fig:sim-wild-mac-thp-4}
    \end{subfigure}
    \hfill
    \begin{subfigure}[t]{0.245\linewidth}
	\centering
	\includegraphics[width=\textwidth]{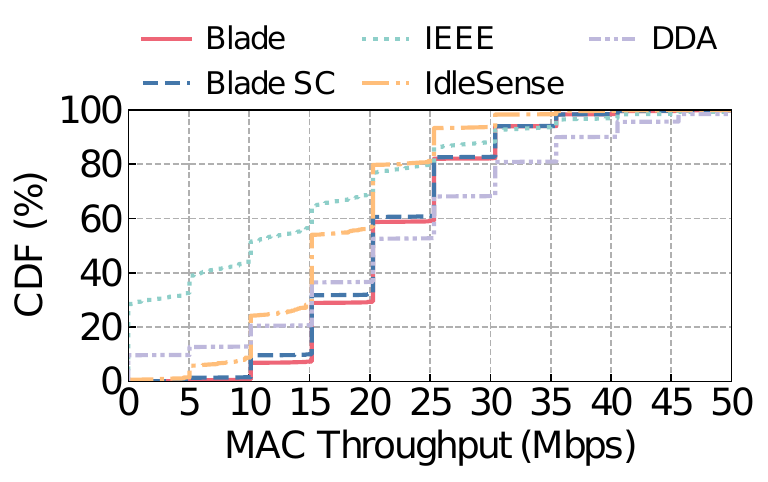}
	\caption{$N = 8$.}
	\label{fig:sim-wild-mac-thp-8}
    \end{subfigure}
    \hfill
    \begin{subfigure}[t]{0.245\linewidth}
	\centering
	\includegraphics[width=\textwidth]{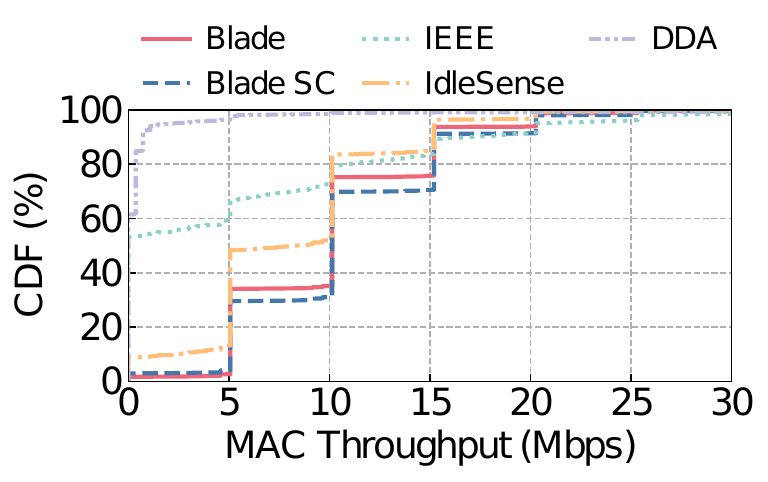}
	\caption{$N = 16$.}
	\label{fig:sim-wild-mac-thp-16}
    \end{subfigure}
    \caption{Distribution of MAC throughput within 100 ms interval under $N$ competing flows.}
    \label{fig:sim-wild-mac-thp}
    \vspace{-2mm}
\end{figure*}

\subsection{Trace-driven Simulation}
We use ns3 for our experimental environment because it accurately simulates the CSMA/CA behavior of Wi-Fi networks \cite{baldo2010-ns3validation, ns3-bianchi} and allows easy modification of the contention control policy. For PHY transmission rate selection, we use Minstrel \cite{mcgregor2010-minstrel}, the default rate adaptation algorithm in both ns3 and the \texttt{mac80211} module of Linux kernel.

\nosection{Baselines.}
We evaluate \sys against the following contention window control mechanisms:
\begin{Itemize}
    \item \textbf{\sys SC}: \sys with only stable-state control logic (\ie HIMD) to demonstrate the effectiveness of the fast recovery policy for collisions;
    \item \textbf{IEEE}:  The default policy in the IEEE 802.11 standard, as explained in \secref{sec:wifi-limit}, using the BE (Best Effort) AC queue ($CW_{min} = 15, CW_{max} = 1023$);
    \item \textbf{IdleSense} \cite{heusse05-idlesense}: It observes the mean number of idle slots between transmission attempts to control the contention window. We provide the transmitter number $N$ to it as it requires such information to operate; 
    \item \textbf{DDA} \cite{yang06-dda}: It controls the contention window to match the backoff delay threshold $\Delta$ imposed by applications. We $\Delta$ to be 5 ms (99th percentile value in \figref{fig:delay-component-indoor}).
\end{Itemize}

\begin{figure*}[t]
\centering
    \begin{minipage}[t]{0.23\linewidth}
	\centering
	\includegraphics[width=\linewidth]{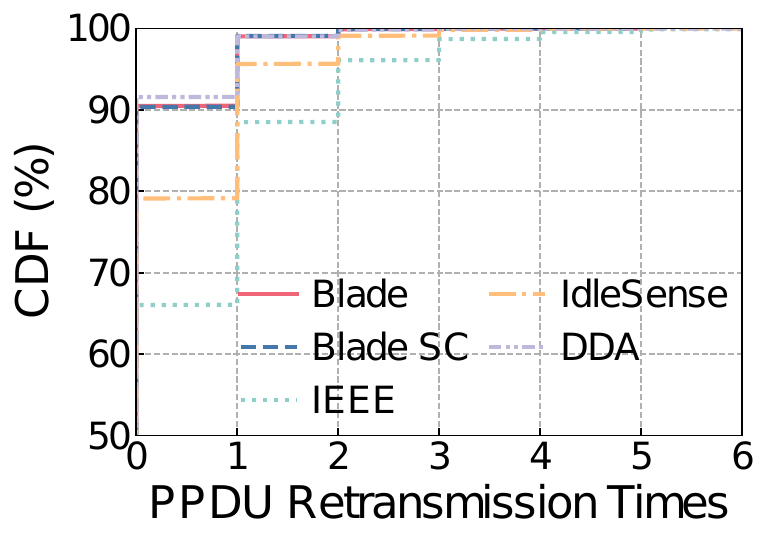}
	\caption{Retransmission times for each PPDU under 8 competing flows.}
	\label{fig:sim-wild-tx-num}
    \end{minipage}
    \hfill
    \begin{minipage}[t]{0.76\linewidth}
	\centering
        \begin{subfigure}[t]{0.49\linewidth}
	\includegraphics[width=\linewidth]{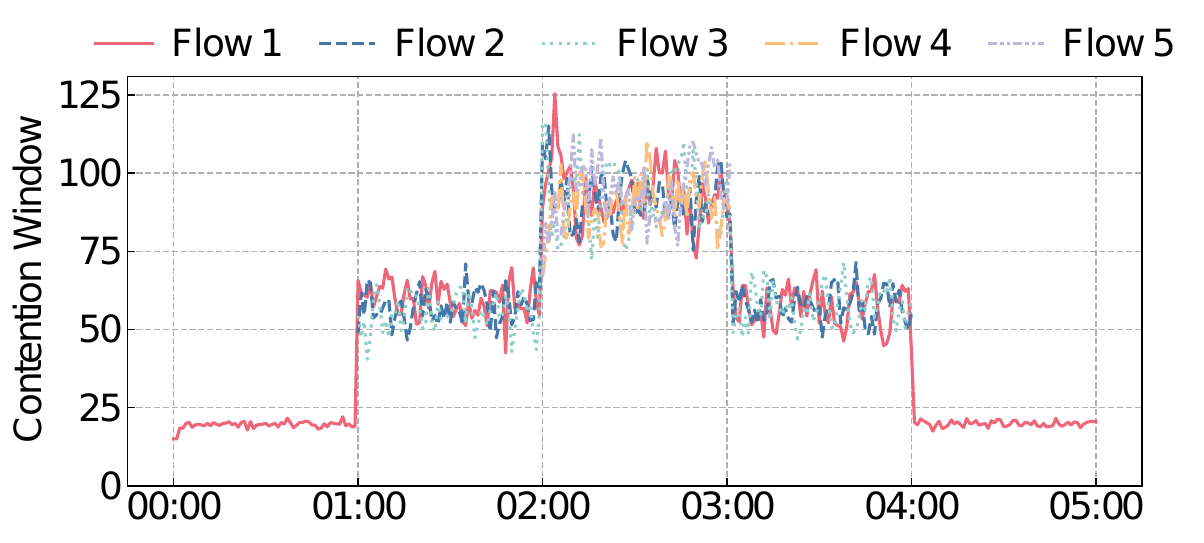}
	\caption{Contention window value.}
	\label{fig:sim-conv-cw}
        \end{subfigure}
        \hfill
        \begin{subfigure}[t]{0.49\linewidth}
	\includegraphics[width=\linewidth]{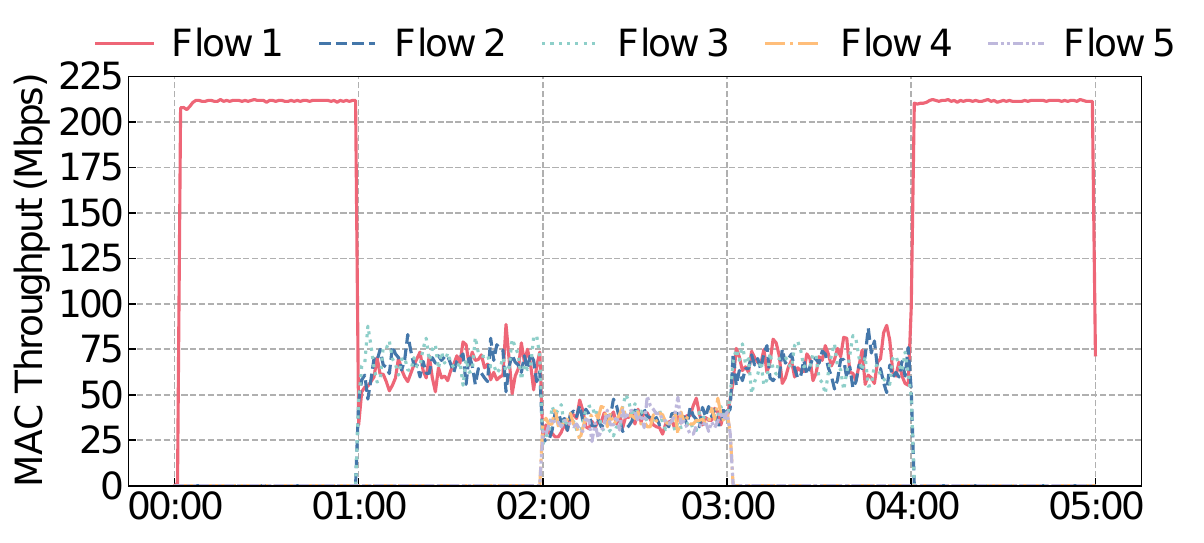}
	\caption{MAC throughput.}
	\label{fig:sim-conv-thp}
        \end{subfigure}
        \caption{Convergence of \sys with five competing flows.}
        \label{fig:wild-convergence}
    \end{minipage}
    \vspace{-2mm}
\end{figure*}

\subsubsection{Saturated Link}\label{sec:eval-wild}
% We first evaluate \sys in an environment with high contention. We deploy five AP-STA pairs using the same channel, each transmitting traffic from AP to STA using the cloud gaming trace we collected. The traffic has a bitrate of approximately 50 Mbps and 60 FPS and lasts 1000 seconds. This experiment evaluates the performance of \sys under intensive contention.

\nosection{Experimental Setup.}
To evaluate the performance of \sys under intensive contention, we deploy $N$ AP-STA pairs ($N = 2, 4, 8, 16$), each transmitting traffic from AP to STA using \texttt{iperf} to saturate the link. The evaluation utilizes the 802.11ax standard (Wi-Fi 6) operating in the 5 GHz band with a 40 MHz bandwidth. All transmitters share the same channel and can hear each other with equal signal strength.

%{\textbf{Why focus on multi-AP contention?} Cloud gaming is downlink-dominated, and in dense residential deployments the dominant latency pathologies come from cross-BSS contention among multiple independently managed APs. In contrast, multiple clients associated to a single AP are largely governed by that AP's internal scheduling/aggregation (e.g., MU features) and do not eliminate cross-BSS collisions and backoff. Blade specifically targets the shared-medium contention across APs.}

\nosection{AP Transmission Latency.}
To demonstrate \sys's effectiveness in reducing tail latency for Wi-Fi last hop, we first evaluate the PPDU transmission latency (\ie frame exchange sequence duration in
\figref{fig:link-latency-decompose}) to show how long a PPDU blocks the AP sending queue.
As shown in \figref{fig:sim-wild-hol}, as the number of competing flows increases from 2 to 16,
the median latency remains similar across all methods.
However, the tail latency increases rapidly for the IEEE 802.11 standard contention control policy,
exceeding 300 ms at the 99th percentile with 8 competing flows.
In contrast, \sys achieves the lowest tail latency among all methods,
limiting the 99.99th percentile latency to 200 ms even with 16 competing flows.
Notably, under 16 competing flows and the standard contention control policy,
we observe frequent AP-STA disconnections due to Beacon
frames experiencing excessively long contention intervals before transmission.
This indicates that standard contention control policy fails
to operate effectively under such high contention levels.
Additionally, \sys without the fast recovery policy shows a slight increase in tail latency,
highlighting the effectiveness of \sys's fast recovery mechanism.

\begin{figure*}[t]
\centering
    \begin{minipage}[t]{0.33\linewidth}
	\centering
	\includegraphics[width=\linewidth]{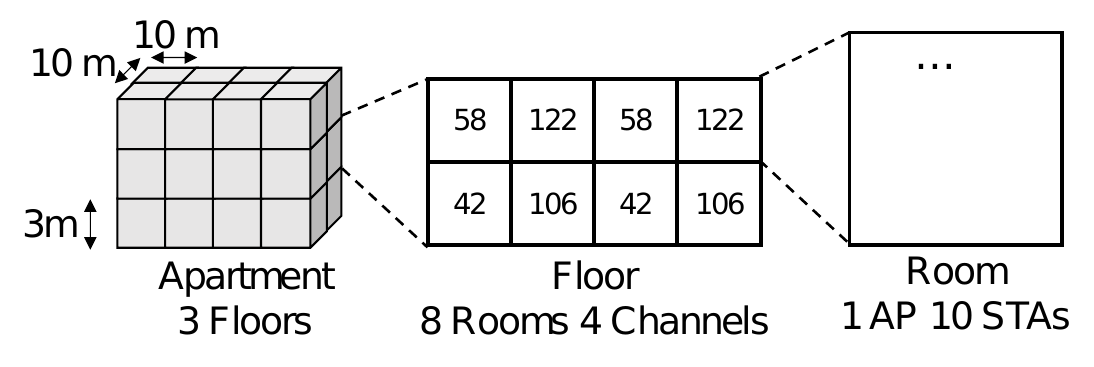}
	\caption{Simulation topology of an apartment.}
	\label{fig:resi-topo}
    \end{minipage}
    \hfill
    \begin{minipage}[t]{0.33\linewidth}
        \centering
	\includegraphics[width=\linewidth]{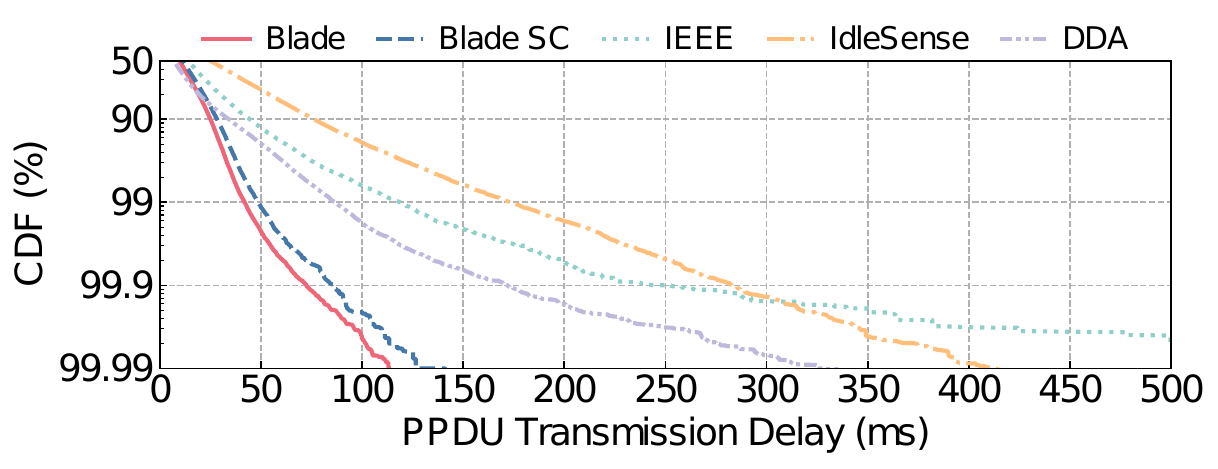}
	\caption{Cloud gaming flow PPDU transmission delay distribution.}
	\label{fig:large-hol}
    \end{minipage}
    \hfill
    \begin{minipage}[t]{0.33\linewidth}
        \centering
	\includegraphics[width=\linewidth]{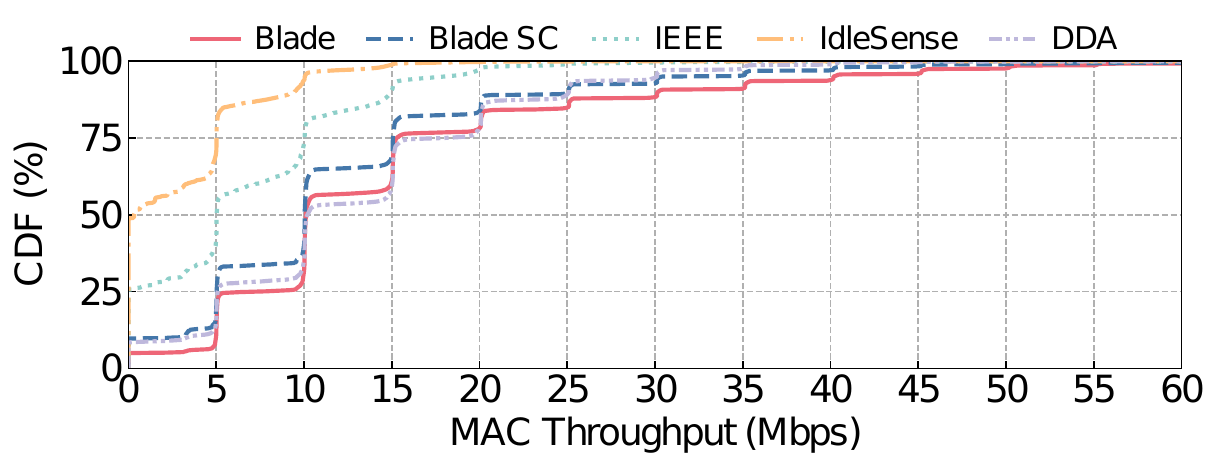}
	\caption{Cloud gaming flow MAC throughput within 100 ms interval.}
	\label{fig:large-mac-thp}
    \end{minipage}
    \vspace{-2mm}
\end{figure*}

\nosection{Retransmission Rate.}
To show \sys's effectiveness in avoiding collisions and improving transmission efficiency,
we plot the distribution of PPDU retransmission counts under
8 competing flows ($N = 8$) in \figref{fig:sim-wild-tx-num}.
Thanks to the stable state control policy,
\sys adapts the contention windows of all transmitters to the channel contention level,
achieving a low retransmission rate.
Specifically, only 10\% of PPDUs are retransmitted once, 
and 1\% are retransmitted twice.
In contrast, under intensive contention,
the standard contention control policy results in 34\% of PPDUs being retransmitted at least once,
with 4\% retransmitted more than twice.

\nosection{MAC Throughput.}
We calculate the MAC throughput in 100 ms
intervals and show the distribution in \figref{fig:sim-wild-mac-thp}.
Due to its lower PPDU retransmission rate and transmission latency,
\sys achieves higher median throughput compared to the standard
policy as the number of competing flows increases.
This result demonstrates that \sys improves MAC layer transmission efficiency for Wi-Fi APs.
Furthermore, \sys results in a steadier and more converged throughput distribution.
In contrast to the standard policy, \sys prevents transient starvation,
where the MAC throughput within 100 ms drops to zero,
demonstrating that \sys achieves fairer bandwidth allocation 
for all transmitters at the micro level.

% \nosection{Fairness.}
% bandwidth max-min per 100/200 ms, CDF

\nosection{Convergence \& Fairness.}
To demonstrate the convergence of \sys, we deploy five AP-STA pairs ($N = 5$) and sequentially start and stop their transmissions over a 5-minute period. As shown in \figref{fig:sim-conv-cw}, with the arrival and departure of competing flows, the contention windows of all transmitters adapt dynamically to the contention level and converge within 1 second. Consequently, \sys quickly achieves a fair bandwidth share among all transmitters, as illustrated in \figref{fig:sim-conv-thp}.

\subsubsection{Real-world Traffic}\label{sec:eval-large}

\nosection{Experimental Setup.}
To evaluate \sys's performance under real-world network traffic, we follow the simulation guidelines outlined in the IEEE standard \cite{ieee-simu-guide} and simulate a three-floor apartment in ns3, as shown in \figref{fig:resi-topo}. Each floor has eight rooms, each with one Wi-Fi AP (central-placed) and ten randomly distributed STAs forming a BSS. In every BSS, the AP sends two cloud gaming flows to two STAs, and the other STAs run real-world traffic trace(video streaming, web browsing, file transfer, etc.).  We utilize four channels (\ie channel numbers 42, 58, 106, and 122) in the 5 GHz band with an 80 MHz bandwidth, ensuring that BSSes in adjacent rooms operate on different channels.

\nosection{Traces.}
We use real-world open-source traces collected from routers \cite{trace-lincolnlab} and base stations \cite{trace-5g}, covering traffic patterns used in our simulation. These traces include timestamps and packet sizes for packet arrivals in both downlink and uplink, representing traffic patterns at wireless last hops. For cloud gaming, we additionally access our cloud gaming platform and collect traffic traces directly from the Wi-Fi router.

\nosection{Performance.}
Following the calculation in \secref{sec:eval-wild},
we plot the PPDU transmission latency and MAC throughput
for cloud gaming flows in \figref{fig:large-hol} and \figref{fig:large-mac-thp}.
With contention from real-world competing network traffic,
\sys constrains the 99.9th and 99.99th percentile latency to 75 ms and 120 ms, respectively.In contrast,
other methods inflate the tail latency to over 300 ms at the 99.99th percentile,
while the standard control policy exceeds 500 ms.
As a result,
\sys achieves only a 5\% starvation rate 
(\ie MAC throughput within 100 ms drops to zero) across all methods,
compared to the 25\% starvation rate observed with the standard control policy.
Notably, DDA and IdleSense perform worse than in the saturated link scenario
because they assume \textit{i.i.d.} traffic patterns from all competing flows,
which is not true in real-world traffic.

\subsection{Microbenchmarks}

\begin{figure}
    \centering
    \begin{subfigure}[t]{0.49\linewidth}
        \includegraphics[width=\linewidth]{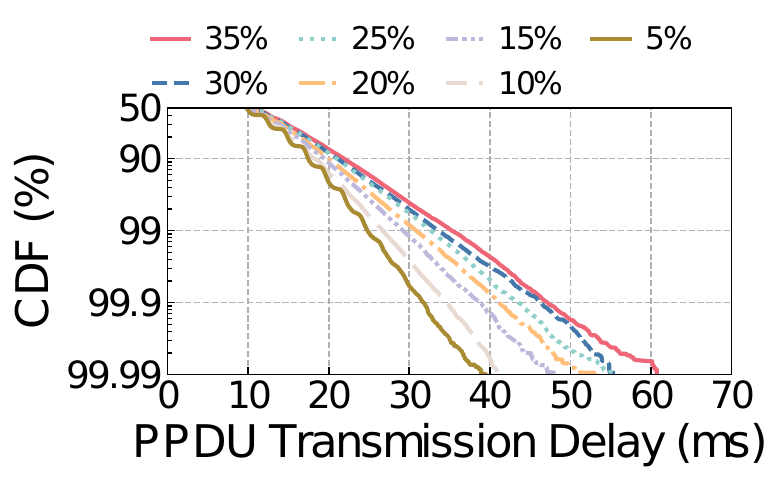}
        \vspace{-4mm}
        \caption{PPDU transmission delay.}
        \label{fig:micro-u-hol}        
    \end{subfigure}
    \hfill
    \begin{subfigure}[t]{0.49\linewidth}
        \includegraphics[width=\linewidth]{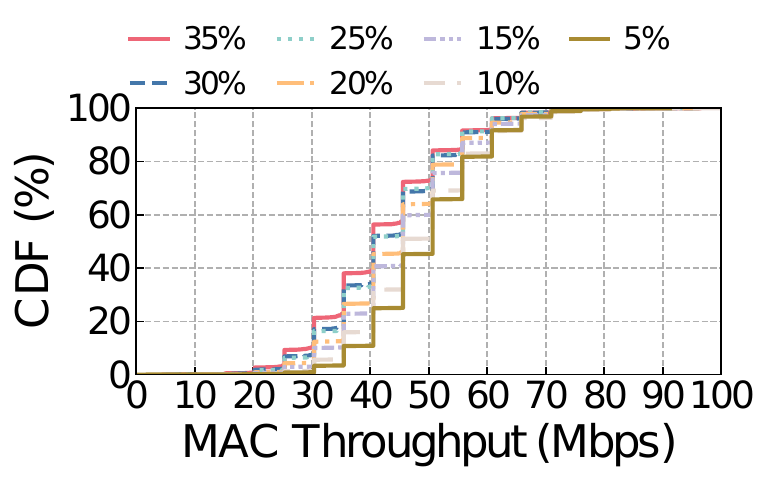}
        \vspace{-4mm}
        \caption{MAC throughput.}
        \label{fig:micro-u-thp}        
    \end{subfigure}
    \caption{Performance of \sys under different target utilization rate $MAR_{tar}$.}
    \label{fig:micro-u}
    % \vspace{-4mm}
\end{figure}

\subsubsection{Influence of Target MAR}\label{sec:microbench-util-rate}
We evaluate the impact of the target MAR on the performance of \sys. We repeat the experiment in \secref{sec:eval-wild} with $N = 4$ and $MAR_{tar}$ varying from 0.05 to $MAR_{max} = 0.35$. As shown in \figref{fig:micro-u}, when $MAR_{tar}$ deviates from the default value of 0.1 within $\pm 0.05$, the performance of \sys remains relatively stable, with a $\pm 5$ ms tail PPDU transmission delay and a $\pm 2.5$ Mbps median MAC throughput deviation. However, as $MAR_{tar}$ approaches $MAR_{max}$, the tail latency increases rapidly, reaching 150\% of the default value. These results align with our analysis in \secref{appendix:design-theory}, which shows that $MAR_{tar} = 0.1$ is an appropriate and robust default value.

% \subsubsection{Effectiveness of Transient Control Policy}

% \subsubsection{Coexistence with IEEE 802.11 Contention Control}

\subsubsection{Parameter Sensitivity}\label{sec:param-sensitivity}
\sys is robust to parameter choices; varying $M_{inc}$, $M_{dec}$, $A_{inc}$, and $A_{fail}$ yields negligible changes in throughput and 

\begin{table}[htbp]
\centering
\scriptsize
\caption{Mobile gaming packet latency distribution (\%)}
\label{tab:rtt_distribution}
\begin{tabularx}{\linewidth}{p{1.0cm}|X|X|X|X|X|X|X|X} % 第一列宽度从1.2cm改为1.0cm
\hline
\multirow{2}{*}{\makecell[c]{RTT\\(ms)}} % 将RTT (ms) 换行以节省空间
& \multicolumn{2}{c|}{\makecell[c]{0 Competing\\Flow}} 
& \multicolumn{2}{c|}{\makecell[c]{1 Competing\\Flow}} 
& \multicolumn{2}{c|}{\makecell[c]{2 Competing\\Flows}} 
& \multicolumn{2}{c}{\makecell[c]{3 Competing\\Flows}}\\
\cline{2-9}
& IEEE & Blade & IEEE & Blade & IEEE & Blade & IEEE & Blade \\
\hline
{[0, 10)}   & 99.7 & 99.8 & 12.4 & 88.6 &  2.1 & 85.9 &  2.3 & 84.1 \\
{[10, 20)}  &  0.3 &  0.2 & 32.1 & 11.2 & 30.6 & 13.8 & 22.7 & 15.7 \\
{[20, 30)}  &  0.0 &  0.0 & 28.3 &  0.2 & 29.5 &  0.1 & 27.8 &  0.1 \\
{[30, 40)}  &  0.0 &  0.0 & 18.1 &  0.0 & 19.0 &  0.2 & 22.7 &  0.1 \\
{[40, 50)}  &  0.0 &  0.0 &  5.3 &  0.0 & 10.1 &  0.0 & 11.3 &  0.0 \\
{[50,100)}  &  0.0 &  0.0 &  3.8 &  0.0 &  8.7 &  0.0 & 13.2 &  0.0 \\
\hline
\end{tabularx}
\vspace{-2mm}
\end{table}

PPDU TX-delay percentiles (details in \secref{append:param-sensitivity}).

\begin{figure*}[t]
\centering
    \begin{minipage}[t]{0.32\linewidth}
	\centering
	\includegraphics[width=\linewidth]{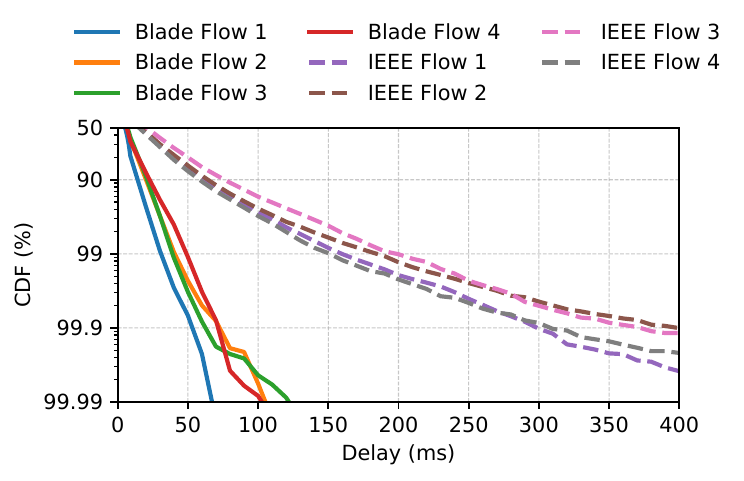}
	\caption{Distribution of transmission delay for four \texttt{iperf} flows.}
	\label{fig:HoLtime_CDF}
    \end{minipage}
    \hfill
    \begin{minipage}[t]{0.32\linewidth}
        \centering
	\includegraphics[width=\linewidth]{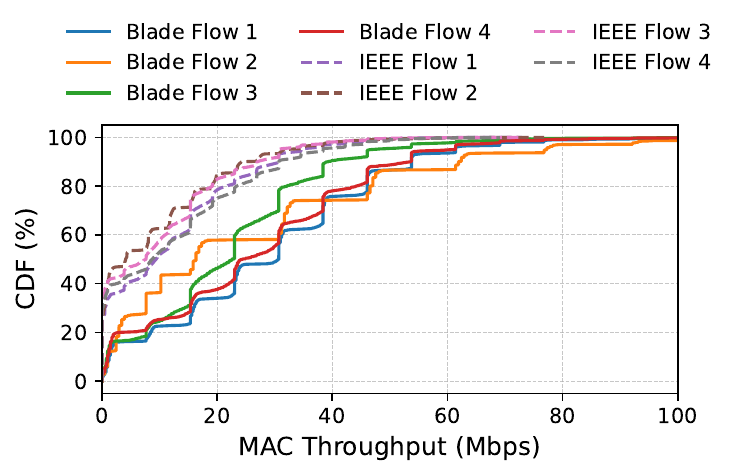}
	\caption{Distribution of MAC throughput for four \texttt{iperf} flows.}
	\label{fig:bw_CDF}
    \end{minipage}
    \hfill
    \begin{minipage}[t]{0.32\linewidth}
        \centering
	\includegraphics[width=\linewidth]{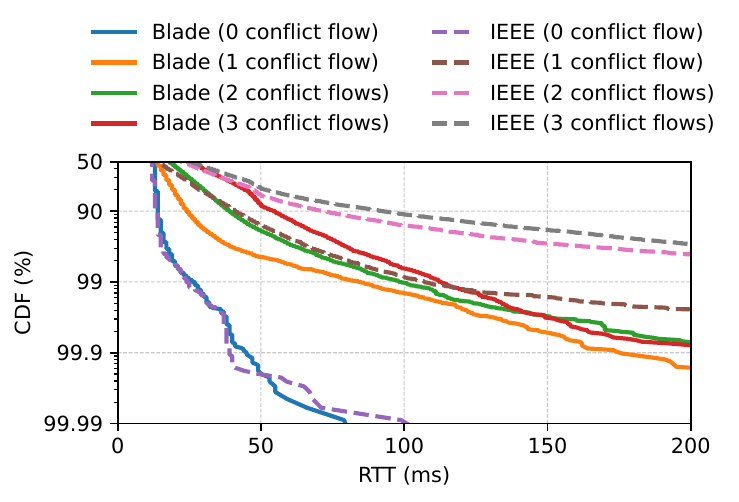}
	\caption{End-to-end frame delay under varying number of \texttt{iperf} flows. }
	\label{fig:START_RTT_CDF}
    \end{minipage}
    \vspace{-2mm}
\end{figure*}

\subsection{Real-World Experiments}\label{Real-World}
% The evaluation includes iPerf-based throughput and delay measurements, cloud gaming latency analysis, and mobile game and file download performance under varying contention levels.

\nosection{Experimental Setup.}
To evaluate the performance of \sys in real world, we conduct experiments using commercial Wi-Fi APs with \sys implemented. We deploy 4 AP-STA pairs. The evaluation utilizes the 802.11ax standard (Wi-Fi 6) operating in the 5 GHz band with 40 MHz bandwidth. All transmitters share the same channel and can hear each other.
\subsubsection{Saturated Links}
We first saturate the wireless link with 4 AP each transmitting an \texttt{iperf} flow to the STA. 
% Four concurrent 100\,Mbps iPerf flows are transmitted over the wireless network, with each flow following either \sys or the IEEE standard CW algorithm. The experiment measures queueing delay and actual wireless bandwidth.
As shown in Fig.~\ref{fig:HoLtime_CDF}, \sys consistently achieves lower tail PPDU transmission delay compared to the IEEE standard, with more than $4\times$ reduction. Therefore, as shown in Fig.~\ref{fig:bw_CDF}, \sys achieves more stable and higher MAC bandwidth utilization than the IEEE standard, demonstrating better adaptation to real-world wireless channel dynamics and reducing inefficiencies caused by excessive contention window growth.

%These results confirm that \sys improves channel efficiency and reduces transmission latency in real router-based experiments, making it more effective for high-throughput scenarios.

\subsubsection{Cloud Gaming}
{We replay our cloud gaming session while injecting 0--3 contending \texttt{iperf} flows. Fig.~\ref{fig:START_RTT_CDF} shows that \sys keeps the 99th-percentile end-to-end frame delay below 100\,ms under heavy contention (vs. $>200\,ms$ for IEEE), cutting stall rate by $>90\%$.} {This directly translates to smoother interactive gameplay, aligning with the motivation in \S\ref{sec:online-measurement}.}

\subsubsection{Mobile Game Traffic}

\noindent we compares the RTT distribution of a mobile game under different numbers of competing flows using the IEEE 802.11 standard and \sys. all flows adopt the same CW adjustment algorithm during the experiment. As shown in \tabref{tab:rtt_distribution},without competing flows, both strategies achieve ultra-low latency. However, IEEE 802.11 performance degrades sharply with more competing flows, with far fewer low-latency packets and higher RTTs. In contrast, \sys maintains over 84\% of packets within {[0,10) ms} even under three competing flows, while IEEE 802.11 drops to only 2.3\%. This demonstrates \sys's effectiveness in mitigating contention and ensuring low-latency performance for mobile gaming applications.

\subsubsection{File Downloading}

\tabref{tab:speed_distribution} presents the speed distribution under different contention levels while downloading a large file, comparing IEEE 802.11 with \sys. Without competing flows, both schemes maintain speeds above 50 Mbps. However, IEEE degrades with 40\% of traffic below 50 Mbps under one competing flow (\sys keeps 94\% at 10–50 Mbps). 
Under heavy contention, 50\% of IEEE traffic drops below 10 Mbps,while 67\% \sys traffic exceeds 20 Mbps. 
These results proving BLADE mitigates contention-induced degradation for more stable throughput.

%\subsubsection{Other Applications}
%We further evaluate \sys with mobile online gaming and file-downloading applications. For online gaming, as shown in Table~\ref{tab:rtt_distribution}, \sys can maintain 84.1\% of end-to-end packet latency within 10ms, compared to 2.3\% in IEEE standard control policy; For file-downloading, as shown in Table~\ref{tab:speed_distribution}, \sys can greatly improve download bandwidth under heavy contention.
\begin{table}[htbp]
\centering
\footnotesize
\caption{Download bandwidth distribution under different contention levels (\%)}
\label{tab:speed_distribution}
\begin{tabularx}{\linewidth}{p{1.0cm}|*{8}{>{\centering\arraybackslash}X}}
\hline
\multirow{2}{*}{\makecell[c]{Bandwidth\\(Mbps)}} 
& \multicolumn{2}{c|}{\makecell[c]{0 Flow}} 
& \multicolumn{2}{c|}{\makecell[c]{1 Flow}} 
& \multicolumn{2}{c|}{\makecell[c]{2 Flows}} 
& \multicolumn{2}{c}{\makecell[c]{3 Flows}}\\
\cline{2-9}
& IEEE & Blade & IEEE & Blade & IEEE & Blade & IEEE & Blade \\
\hline
0--5      & 0   & 0   & 1   & 0   & 0   & 0   & 1   & 0 \\
5--10     & 0   & 0   & 5   & 0   & 43  & 1   & 79  & 0 \\
10--20    & 0   & 0   & 50  & 2   & 57  & 17  & 10  & 24 \\
20--30    & 0   & 0   & 3   & 3   & 0   & 71  & 0   & 74 \\
30--40    & 0   & 0   & 0   & 52  & 0   & 9   & 0   & 2 \\
40+       & 100 & 100 & 41  & 43  & 0   & 2   & 0   & 0 \\
\hline
\end{tabularx}
\vspace{-2mm}
\end{table}
Overall, these real-world experiments demonstrate that \sys effectively optimizes CW adjustment, accommodating both throughput and latency-sensitive applications. 
%Unlike the IEEE standard, \sys dynamically adapts to network conditions, ensuring better performance in practical deployment scenarios.

%% file: sections/70-discussion.tex
\section{Discussion}

\nosection{Why not centralized scheduling?}
%{A natural alternative to contention control is to avoid contention entirely via explicit scheduling (\eg TDMA-style airtime reservation or a centralized controller that assigns transmission turns). While effective in managed deployments, these approaches typically require tight coordination (time synchronization and/or control-plane messaging), changes to the MAC protocol, and often a common administrative domain. In contrast, \sys intentionally stays within the CSMA/CA framework and only changes local CW adaptation, enabling a fully distributed design that can be deployed incrementally on commodity APs without requiring cooperation from neighboring APs or stations.}
{Explicit scheduling (e.g., TDMA-style airtime reservation) avoids contention, but requires tight coordination and a common administrative domain, which is unachievable for Wi-Fi routers purchased and controlled by end consumers who lack centralized management capabilities.In contrast, \sys operates within CSMA/CA, only adjusting local contention windows (CW). Its fully distributed design enables incremental deployment in commodity APs without neighboring coordination.}

\nosection{Coexistence with IEEE 802.11 Contention Control.}
%\sys adapts its contention window to the observed channel contention level, as it reaches fair convergence when all Wi-Fi transmitters implement \sys, it may unintentionally provide more transmission chances to transmitters using the IEEE standard control policy when operating on the same channel, thus degrading the MAC throughput. This occurs because the standard policy is unaware of the contention level and typically keeps $CW = CW_{min}$ most of the time. We show in \secref{append:coexist} that configuring \sys with a higher {$MAR_{tar}$} value allows it to better compete with the standard control policy.
%{More broadly, \sys is incrementally deployable: a BLADE-enabled transmitter measures MAR over \emph{all} observed transmissions, including legacy devices, and can still suppress extreme contention-driven droughts for its own traffic. Fairness convergence is strongest when all transmitters adopt \sys, while partial deployment primarily improves tail latency for participating transmitters.}
\sys achieves fair convergence when universally deployed, but may inadvertently cede transmission opportunities to IEEE 802.11-compliant devices (which typically retain small contention windows). As shown in \secref{append:coexist},configuring \sys with a higher {$MAR_{tar}$} enhances its competitiveness with legacy devices. Notably, \sys supports incremental deployment: even on partial adoption, it suppresses contention-driven packet-delivery droughts for its own traffic, while full deployment maximizes fairness and latency performance.

\nosection{Hidden Terminal.}
Since \sys relies on the consensus signal from the same channel, it may be affected by the Hidden Terminal Problem \cite{hidden-terminal}, where transmitters perceive different transmission opportunity utilization rates. 
{In this setting, MAR should be interpreted as a ``local'' contention signal within a carrier-sensing domain rather than a globally consistent metric.} RTS/CTS is widely used to mitigate this issue. Since a CTS is followed by a PPDU transmission from a hidden terminal, upon receiving CTS, \sys can infer that two transmission opportunities have been utilized when calculating MAR. We show in \secref{append:hidden-terminal} that \sys maintains low PPDU transmission delay for all transmitters in the presence of hidden terminals.

%\nosection{QoS Priority.}
%\sys is designed to achieve micro-level bandwidth %fairness for all Wi-Fi transmitters. However, it can %also provide priority-based transmission allocation %for applications with different QoS requirements. %Similar to the EDCA mechanism defined in the IEEE %802.11e standard \cite{ieee80211e}, \sys can assign %different application flows to distinct AC queues, %each with its own [$CW_{min}, CW_{max}$] range.

%% file: sections/80-related.tex
\section{Related Work}

\nosection{Real-Time Streaming in Wireless Networks.}
% CC, MP, loss recovery
Many prior studies have discussed the latency bottleneck of wireless networks in real-time streaming. They propose various methods to reduce the tail frame delivery latency, including congestion control algorithms \cite{shibo24-pudica}, multipath transmission \cite{yuhan24-augur}, and novel loss recovery schemes at both transport layer \cite{zili24-hairpin} and application layer \cite{yihua24-grace}. These studies regard wireless fluctuations as inherent to the link and rely on \textit{indirect} solutions from higher layers to alleviate its impact. Orthogonal to all these methods, \sys takes a \textit{direct} approach on the link layer to mitigate the long tail latency induced by Wi-Fi last hop and can be jointly deployed with them.

\nosection{Wi-Fi Performance Enhancement.}
Prior work improves Wi-Fi via rate adaptation, contention control, channel selection, and AQM~\cite{lacage04-aarf, mcgregor2010-minstrel, heusse05-idlesense, yang06-dda, Vasudevan05-ap-selection, toke17-wifi-aqm}. Few focus on latency-sensitive flows~\cite{zili22-zhuge, changhua17-qair}. \sys targets contention-driven tail latency with no traffic-pattern assumptions.

%% file: sections/90-conclusion.tex
\section{Conclusion}
In this paper, we reveal the fundamental limitation of the contention window adjustment mechanism in IEEE 802.11 standard and identify it to be the root cause of long tail video frame delivery latency of next-generation real-time communication applications (NGRTC) in Wi-Fi networks. We present \sys, a novel contention window control algorithm. Compared to the standard, \sys can significantly reduce Wi-Fi transmission latency and improve MAC throughput. We believe \sys to be an important building block towards the rapid development of NGRTC.

%% file: sections/99-apendix.tex
\appendix

\section{PPDU Contention Interval Calculation}\label{append:hol-calc}

\begin{figure*}
\centering
    \begin{subfigure}[t]{\linewidth}
        \centering
        \includegraphics[width=0.8\textwidth]{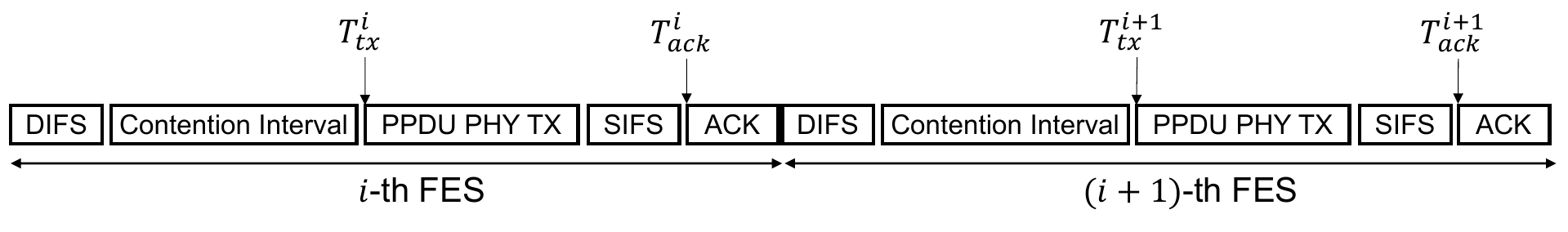}  
        \caption{Transmission success.}
        \label{fig:hol-calc-success}     
    \end{subfigure}
    \begin{subfigure}[t]{\linewidth}
        \centering
        \includegraphics[width=0.8\textwidth]{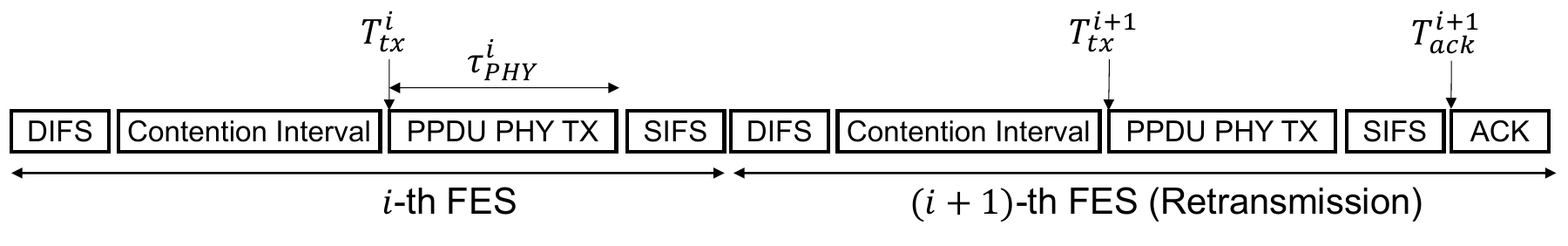}
        \caption{Transmission failure.}
        \label{fig:hol-calc-fail} 
    \end{subfigure}
    \caption{Illustration of PPDU contention interval calculation.}
    \label{fig:hol-calc}
\end{figure*}

Since the random backoff procedure is handled by WNIC firmware, it is not easy to directly acquire the contention interval for each PPDU transmitted in Wi-Fi. Therefore, we adopt a passive approach: we deploy an AP-STA pair and use \texttt{iperf} to keep the AP WNIC busy. We place a Wi-Fi sniffer closely to the AP to capture all traffic (including PPDUs and ACKs) related to the AP. As shown in \figref{fig:hol-calc-success}, because the WNIC keeps busy, the frame exchange sequence (FES) of the $i$-th PPDU is immediately followed by the FES of the $(i + 1)$-th PPDU. From the sniffed trace, we can acquire the precise timestamp of PHY transmission event $T_{tx}^i$ and ACK event $T_{ack}^i$ of the $i$-th PPDU. Since the DIFS, SIFS and ACK are standard intervals with fixed values, we can calculate the contention interval of the $(i + 1)$-th PPDU as $T_{tx}^{i+1} - T_{ack}^i - ACK - DIFS$. The PHY transmission time of the $i$-th PPDU can be calculated as $T_{ack}^i - T_{tx}^i - SIFS$.

Upon transmission failures, no ACK frame is sniffed. As illustrated in \figref{fig:hol-calc-fail}, in this case, we can calculate the PHY transmission time of the $i$-th PPDU $\tau_{PHY}^i$ as we can acquire the PPDU size and transmission rate from the sniffed trace. Therefore, the contention interval of the $(i + 1)$-th PPDU can be calculated as $T_{tx}^{i+1} - T_{tx}^i - \tau_{PHY}^i - SIFS - DIFS$.

\section{Limitation of Priority-Based IEEE 802.11 Contention Control}\label{append:80211-qos-limit}

\begin{figure}[t]
\centering
    \begin{subfigure}[t]{0.49\linewidth}
	\centering
	\includegraphics[width=\textwidth]{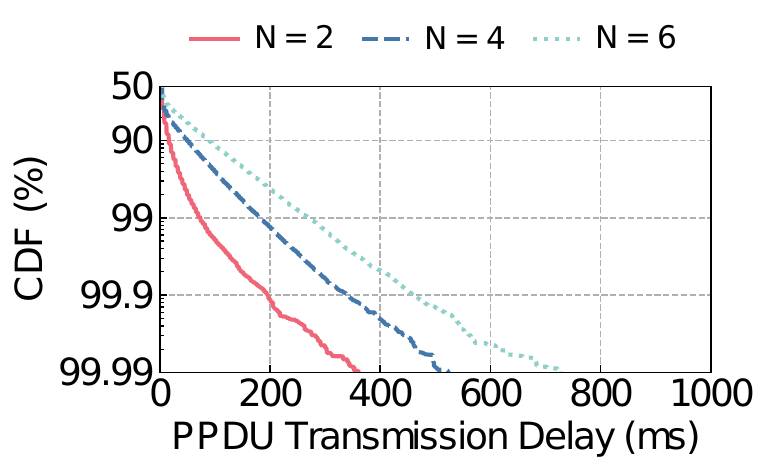}
	\caption{PPDU transmission delay.}
	\label{fig:sim-wild-hol-vi}
    \end{subfigure}
    \hfill
    \begin{subfigure}[t]{0.49\linewidth}
	\centering
	\includegraphics[width=\textwidth]{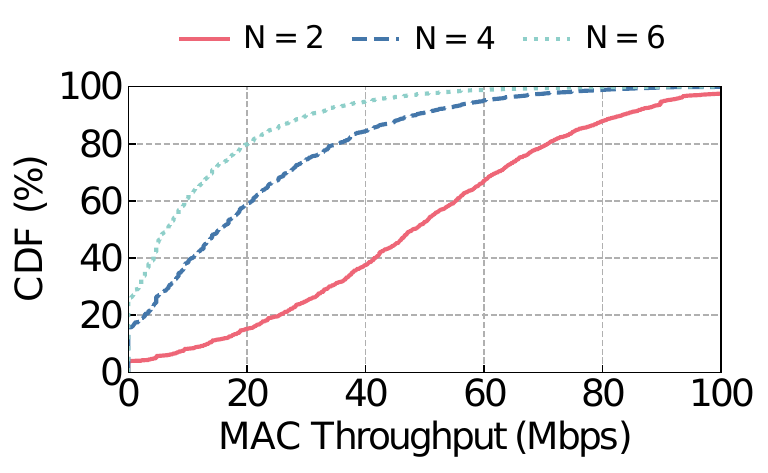}
	\caption{MAC throughput.}
	\label{fig:sim-mac-thp-vi}
    \end{subfigure}
    \caption{Performance of VI AC queue with $N$ competing flows.}
    \label{fig:sim-wild-vi}
\end{figure}

\begin{figure}[t]
    \centering
    \begin{subfigure}[t]{0.49\linewidth}
        \includegraphics[width=\linewidth]{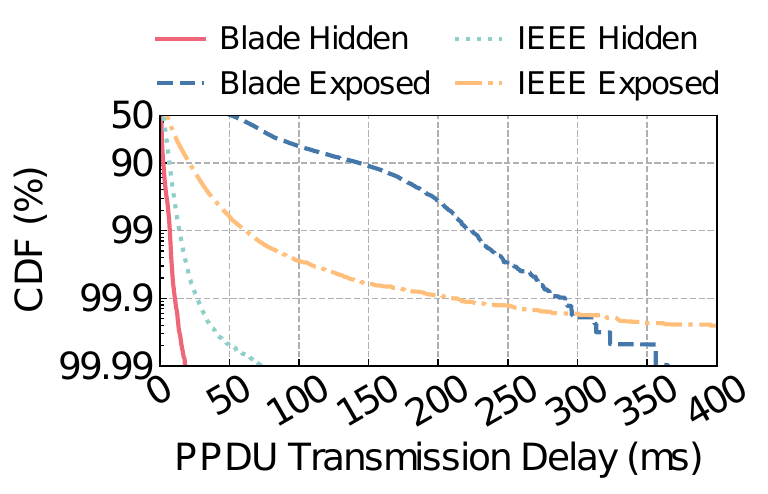}
        \vspace{-4mm}
        \caption{RTS/CTS disabled.}
        \label{fig:micro-hidden-norts}        
    \end{subfigure}
    \hfill
    \begin{subfigure}[t]{0.49\linewidth}
        \includegraphics[width=\linewidth]{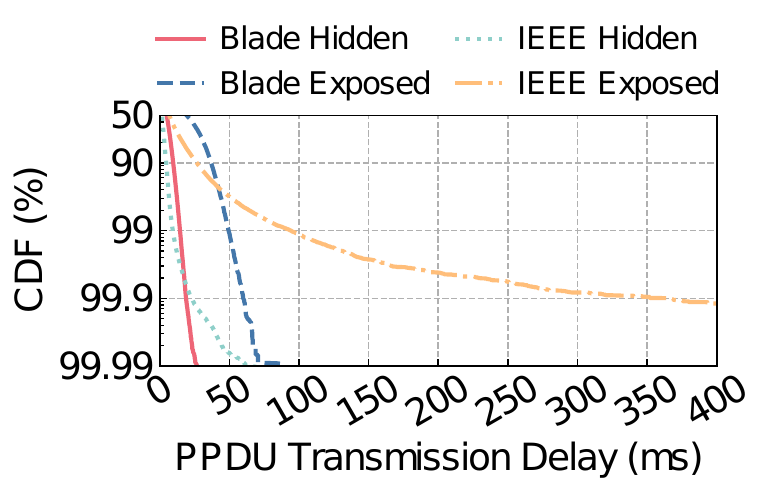}
        \vspace{-4mm}
        \caption{RTS/CTS enabled.}
        \label{fig:micro-hidden-rts}        
    \end{subfigure}
    \caption{Influence of hidden terminal on \sys with RTS/CTS disabled/enabled.}
    \label{fig:micro-hidden}
\end{figure}

\begin{figure*}[t]
    \begin{subfigure}[t]{0.33\linewidth}
	\centering
	\includegraphics[width=\linewidth]{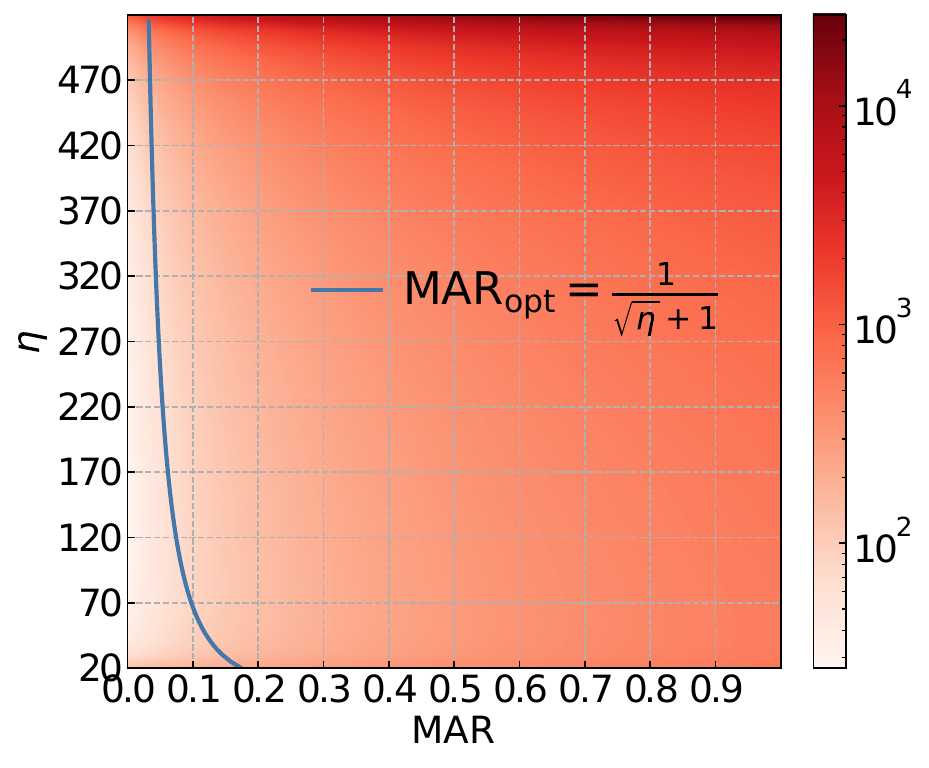}
	\caption{$N = 2$.}
    \end{subfigure}
    \hfill    
    \begin{subfigure}[t]{0.33\linewidth}
	\centering
	\includegraphics[width=\linewidth]{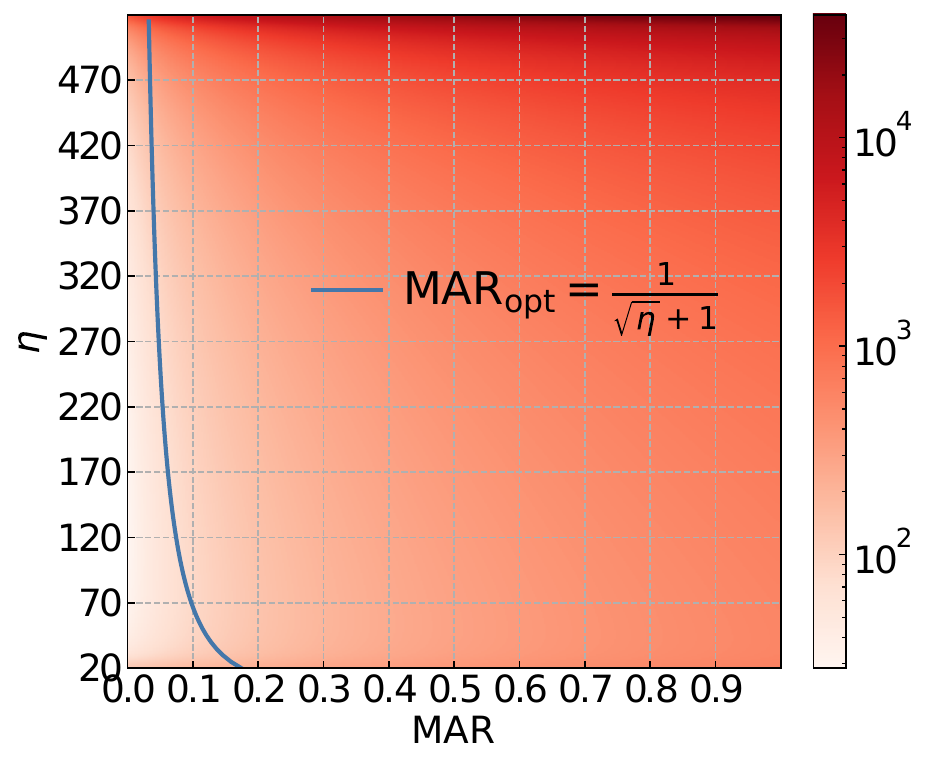}
	\caption{$N = 4$.}
    \end{subfigure}
    \hfill
    \begin{subfigure}[t]{0.33\linewidth}
	\centering
	\includegraphics[width=\linewidth]{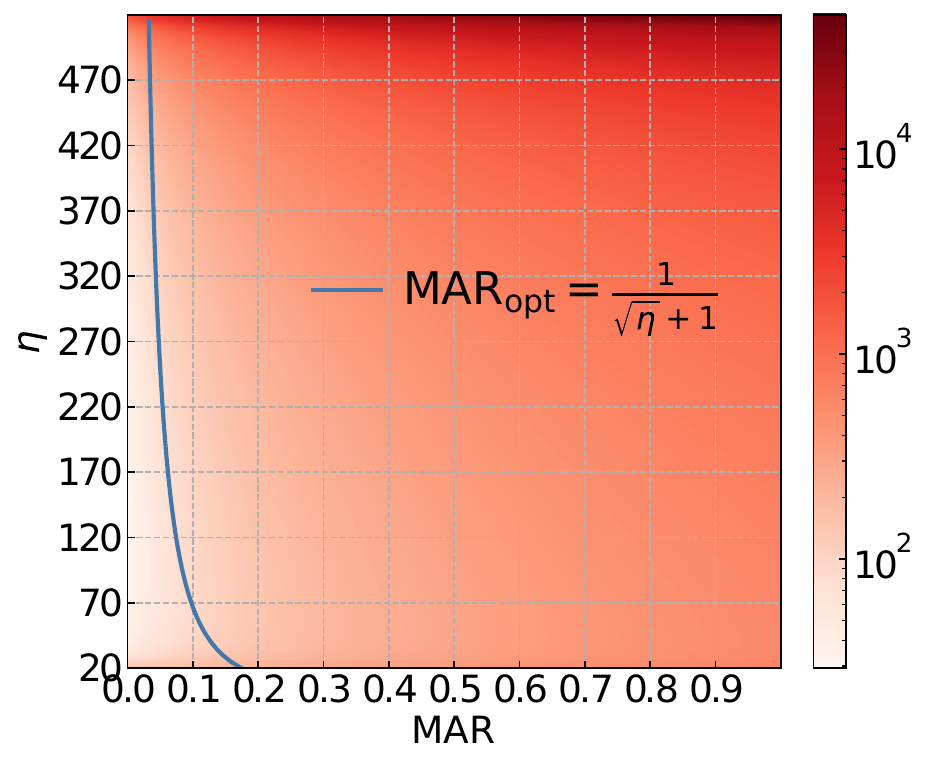}
	\caption{$N = 8$.}
    \end{subfigure}
    \begin{subfigure}[t]{0.33\linewidth}
	\centering
	\includegraphics[width=\linewidth]{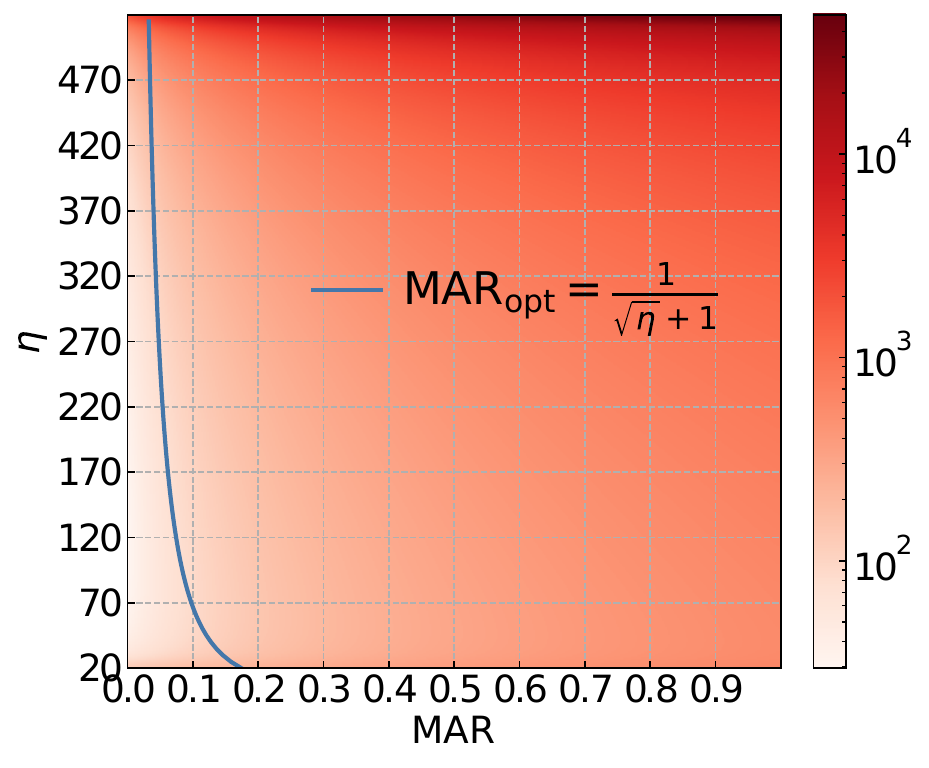}
	\caption{$N = 16$.}
    \end{subfigure}
    \hfill    
    \begin{subfigure}[t]{0.33\linewidth}
	\centering
	\includegraphics[width=\linewidth]{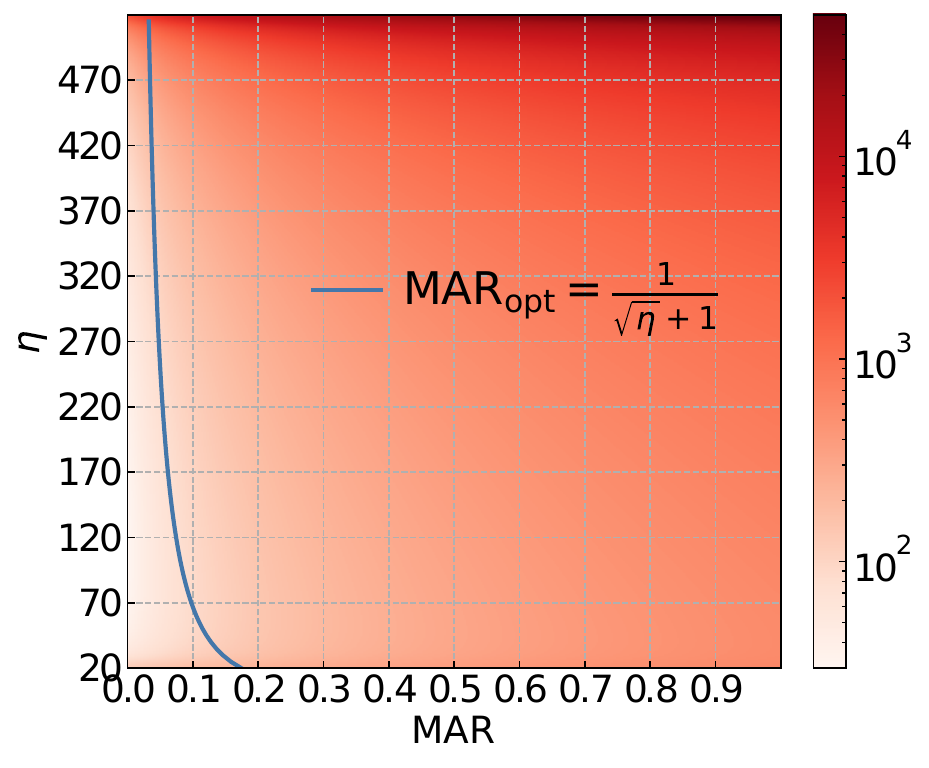}
	\caption{$N = 32$.}
    \end{subfigure}
    \hfill
    \begin{subfigure}[t]{0.33\linewidth}
	\centering
	\includegraphics[width=\linewidth]{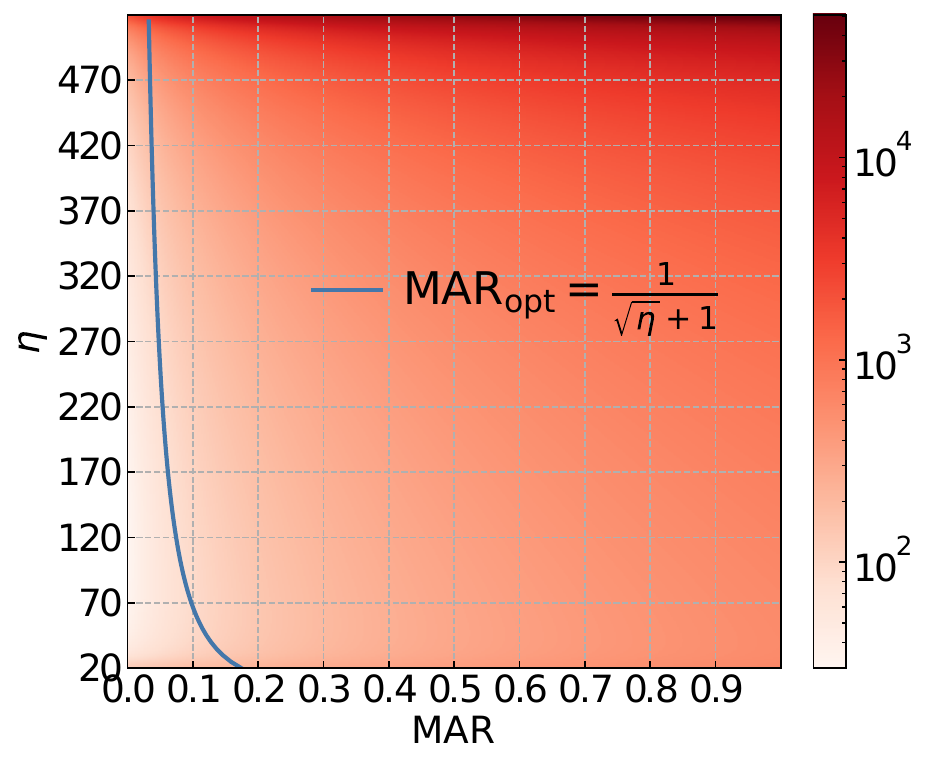}
	\caption{$N = 64$.}
    \end{subfigure}
    \caption{$\mathcal{L}(MAR)$ dynamics with respect to different values of $MAR$ and $\eta$ under different transmitter numbers $N$. The blue line illustrates the optimal $MAR$ value for each $\eta$.}
    \label{fig:eta-mar-L}
\end{figure*}

\begin{figure}
    \centering
    \begin{subfigure}[t]{0.49\linewidth}
        \includegraphics[width=\linewidth]{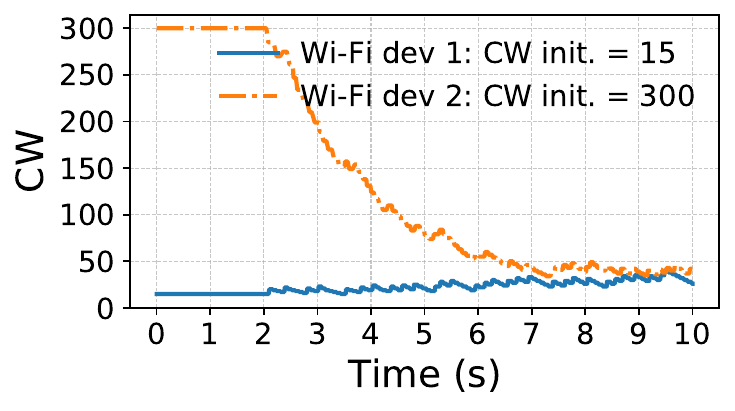}
        \vspace{-4mm}
        \caption{Traditional AIMD.}
    \label{fig:cw-aimd}
    \end{subfigure}
    \hfill
    \begin{subfigure}[t]{0.49\linewidth}
        \includegraphics[width=\linewidth]{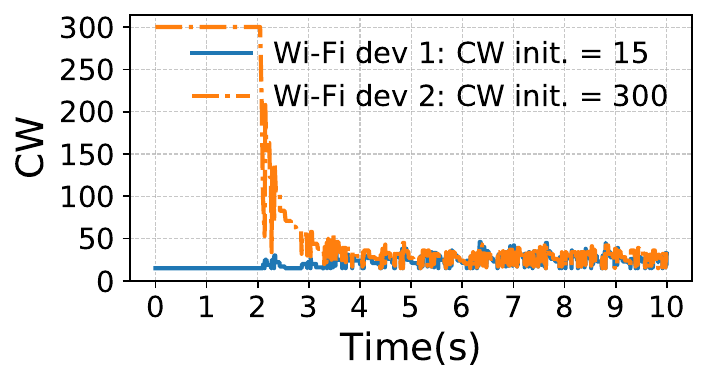}
        \vspace{-4mm}
        \caption{\sys's HIMD}
    \label{fig:cw-proposed}
    \end{subfigure}
    \caption{The comparison of convergence speed of traditional AMID and HIMD of \sys.}
    \label{fig:cw_convergence}
\end{figure}

IEEE 802.11e standard \cite{ieee80211e} has proposed Enhanced Distributed Channel Access (EDCA), defining several Access Categories (ACs) queues with different $CW_{min}$ and $CW_{max}$ values to accommodate various QoS requirements. Specifically, there are BK (Background) queue with $CW_{min} = 7, CW_{max} = 1023$ BE (Best Effort) queue with $CW_{min} = 15, CW_{max} = 1023$, VI (Video) queue with $CW_{min} = 7, CW_{max} = 15$, and VO (Voice) queue with $CW_{min} = 1, CW_{max} = 3$. While the BE queue is the default, adopting AC queues with higher priority can lower the CW value and occupy more transmission chances. However, with the rapid development of RTC applications, it is common to have multiple RTC sessions in the same wireless environment nowadays. When multiple flows with high priority contend for transmission chances in the same channel, the contention level can be severely intensified, leading to more collisions. We demonstrate this by repeating the experiment in \secref{sec:eval-wild}, with $N$ ($N = 2, 4, 6$) \texttt{iperf} flows saturating the link with the VI queue. As shown in \figref{fig:sim-wild-vi}, with competing flows from VI queues, the PPDU transmission delay significantly increases even with $N = 2$, compared to the BE queue (the 99.99th percentile delay is 56 ms for BE queue, as shown in \figref{fig:sim-wild-hol-2}). Therefore, the MAC throughput within 100 ms interval exhibits a more unsteady pattern, with 19\% starvation rate when $N = 4$ (the starvation rate for BE queue when $N = 4$ is 4\%, as shown in \figref{fig:sim-wild-mac-thp-4}).

\section{Additional Evaluation Results}\label{append:extra-eval}
\subsection{Parameter Sensitivity}\label{append:param-sensitivity}
% paste the original table block here

We evaluate the parameter sensitivity of \sys by repeating the experiment in \secref{sec:eval-wild} with $N = 4$ and different parameter values. As shown in \tabref{tab:param-sensitivity}, changes in parameter values lead to negligible performance shifts compared to the default configuration. Therefore, \sys is robust and not sensitive to its parameters.
\vspace{-2mm}
\begin{table}[htb]
\resizebox{\linewidth}{!}{
\begin{tabular}{ccc}
\Xhline{2\arrayrulewidth}
\textbf{Variant} & \textbf{\begin{tabular}[c]{@{}c@{}}Avg. MAC \\ Throughput (Mbps)\end{tabular}} & \textbf{\begin{tabular}[c]{@{}c@{}}50/95/99/99.9/99.99th \\ PPDU TX Delay (ms)\end{tabular}} \\
\Xhline{2\arrayrulewidth}
\sys Default  & 48.5 & 9.8/21.0/26.7/34.6/42.1                                                                          \\
$M_{inc} = 250$  & 48.1  & 9.8/21.2/27.3/35.2/43.1                                                                          \\
$M_{inc} = 125$  & 48.6   & 9.7/21.0/26.6/34.3/42.4                                                                          \\
$M_{dec} = 0.85$ & 48.5   & 9.7/21.4/27.7/36.5/44.9                                                                          \\
$M_{dec} = 0.75$ & 48.1  & 9.7/21.9/28.6/37.8/46.4                                                                          \\
$A_{inc} = 10$   & 48.2  & 9.8/21.2/26.9/35.0/42.1                                                                          \\
$A_{inc} = 30$   & 48.9 & 9.6/21.7/28.2/37.1/46.1                                                                          \\
$A_{fail} = 10$  & 48.6  & 9.7/21.2/27.1/35.3/42.5                                                                          \\
$A_{fail} = 20$  & 48.7 & 9.7/21.7/28.1/37.7/44.4 \\
\Xhline{2\arrayrulewidth}
\end{tabular}
}
\vspace{2mm}
\caption{Parameter Sensitivity of \sys.}
\label{tab:param-sensitivity}
\end{table}

\section{Detailed Anatomy of Packet Delivery Droughts}\label{append:drought-mech}

This appendix expands the mechanism summary in \secref{sec:wifi-limit}. We provide (i) ns-3 simulation evidence that collisions increase retransmissions under multi-AP contention, (ii) controlled AP experiments validating the same effect in practice, and (iii) a packet-level example and delay decomposition illustrating how countdown freezes amplify contention intervals.

\begin{figure}[t]
\centering
    \begin{minipage}[t]{0.49\linewidth}
    \centering
    \includegraphics[width=\textwidth]{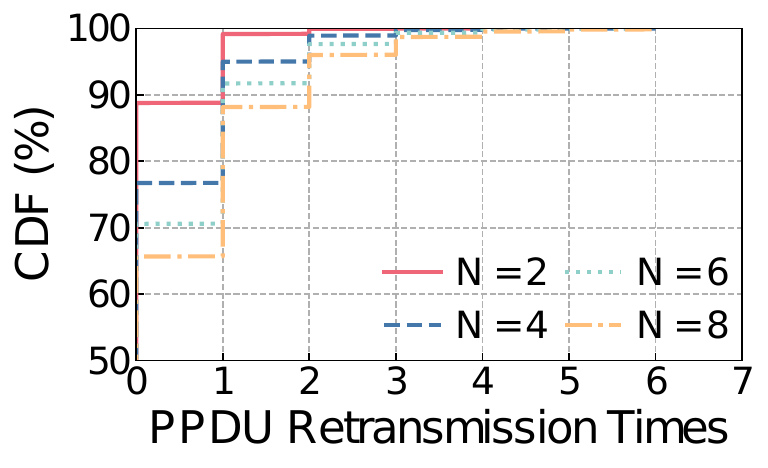}
    \caption{PPDU retransmission times with $N$ competing flows.}
    \label{fig:moti-tx-num}
    \end{minipage}
    \begin{minipage}[t]{0.49\linewidth}
    \centering
    \includegraphics[width=\textwidth]{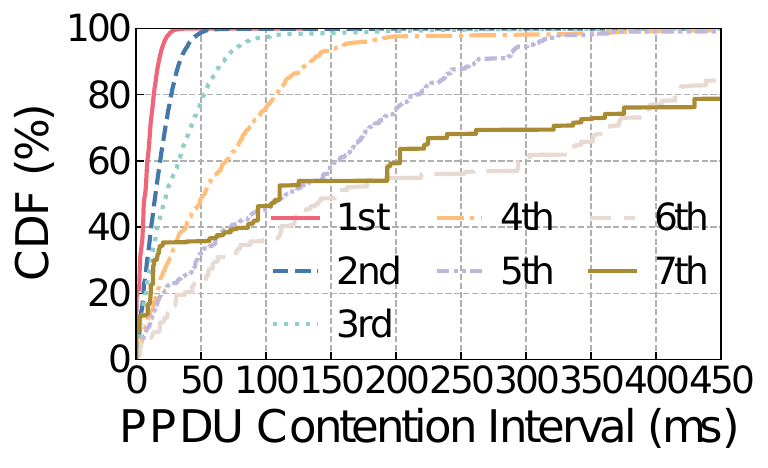}
    \caption{PPDU backoff time at the n-th transmission with $N = 6$.}
    \label{fig:moti-backoff-tx}
    \end{minipage}
    \vspace{-4mm}
\end{figure}

To understand why packets fail to be delivered within measurement intervals, 
we conducted systematic NS3 simulations, which accurately model Wi-Fi's CSMA/CA behavior~\cite{baldo2010-ns3validation, ns3-bianchi}. 
We deploy $N$ Wi-Fi APs contending in the same channel, 
each transmitting \texttt{iperf} flows to saturate the link. 
Our analysis reveals two key factors that lead to packet delivery droughts:

First, packet retransmissions significantly extend the total delivery time, 
as each failed attempt requires additional transmission attempts. 
\figref{fig:moti-tx-num} demonstrates how channel contention directly impacts retransmission frequency: 
with 2 competing devices ($N = 2$), almost all packets are delivered successfully on first attempt. 
However, as contention increases, packets require more retransmission attempts. 
When eight devices compete ($N = 8$), 
34\% of packets need at least one retransmission, 
with some requiring up to 6 retries. 
This means a single packet can require multiple transmission 
attempts before successful delivery, substantially extending its delivery time.

Second, and more critically, 
each retransmission triggers a vicious cycle due to Wi-Fi's exponential backoff mechanism. 
When a transmission fails, the device doubles its contention window, 
effectively reducing its ability to compete for channel access compared to devices with smaller windows. 
\figref{fig:moti-backoff-tx} demonstrates this effect by tracking 
contention intervals across successive retransmission attempts (with $N = 6$). 
While the first transmission attempt experiences relatively short contention intervals, 
each subsequent attempt faces progressively longer delays due to the enlarged contention window. 
By the sixth retransmission, over 60\% of PPDUs experience contention intervals exceeding 200ms.
As a result, the transmission delay for each PPDU (\ie frame exchange sequence duration in \figref{fig:link-latency-decompose}) 
significantly increases with competing flow number $N$, as shown in \figref{fig:moti-hol}. 
This combination of frequent retransmissions and extended contention intervals 
explains the packet delivery droughts we observed in production networks.

\subsubsection{Validating Simulation Results with Wi-Fi AP}
To complement our simulation findings and validate them in real-world conditions, 
we conducted controlled experiments using commercial Wi-Fi APs. 
While our simulations provide insights into the fundamental relationship between contention and packet delivery, 
real-world factors such as channel dynamics and interference could affect these behaviors. 
We set up a testbed using Xiaomi AX3600 Wi-Fi APs in a typical office environment, 
where they experience natural channel contention from existing network traffic. 
Through these experiments, we aim to verify whether the packet delivery patterns and retransmission behaviors observed in simulations manifest similarly in practice.

In our experiments, we established a saturated link between the AP and a client device (STA) using \texttt{iperf}, 
allowing us to observe the system under consistent load. 
By analyzing air-sniffed traces, we calculated precise PHY transmission delay 
and contention interval for each PPDU (detailed methodology in \secref{append:hol-calc}). 
This controlled setting enabled us to dissect the transmission process 
and identify the key factors leading to packet delivery drought.

\begin{figure}[t]
\centering
\begin{minipage}[t]{0.49\linewidth}
    \centering
    \includegraphics[width=\textwidth]{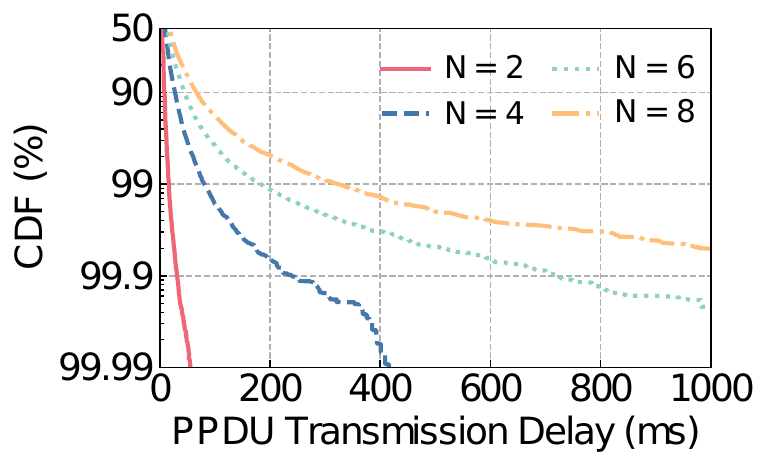}
    \caption{PPDU transmission delay with $N$ competing flows.}
    \label{fig:moti-hol}
    \end{minipage}
    \hfill
    \begin{minipage}[t]{0.49\linewidth}
    \centering
    \includegraphics[width=\textwidth]{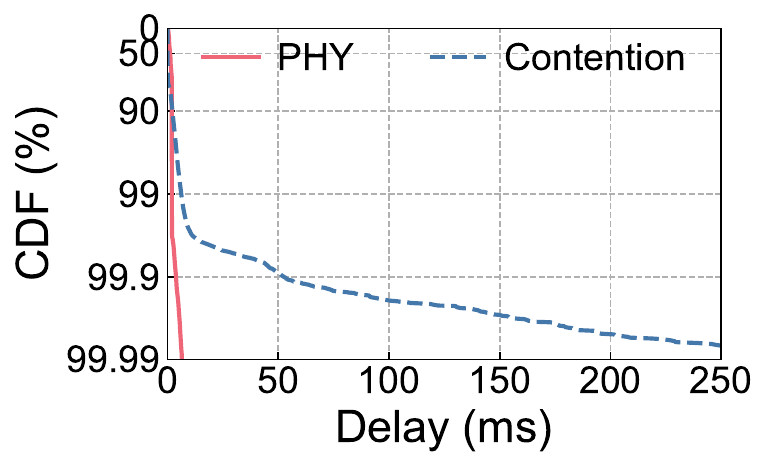}
    \caption{Contention interval and PHY latency distribution for each Wi-Fi PPDU.}
    \label{fig:delay-component-indoor}
    \end{minipage}
\end{figure}

\begin{figure}[t]
    % \centering
    \includegraphics[width=0.49\textwidth]{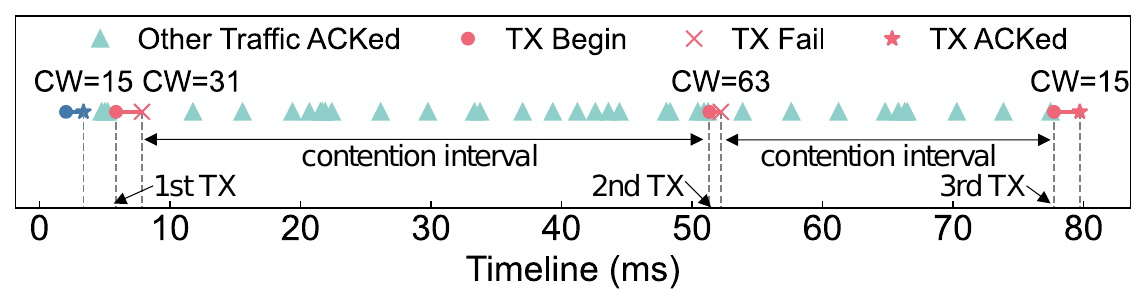}
    \caption{Lifetime of a single PPDU (in red). Green triangles illustrate the competing traffic from other devices.}
    \label{fig:moti-example}
\end{figure}

% \begin{figure}[t]
% 	\centering
% 	\begin{subfigure}[t]{0.49\linewidth}
% 		\centering
% 		\includegraphics[width=\textwidth]{figure/rate-delta-clear.pdf}
% 		\caption{Wilderness.}
% 		\label{fig:rate-delta-clear}
% 	\end{subfigure}
% 	\begin{subfigure}[t]{0.49\linewidth}
% 		\centering
% 		\includegraphics[width=\textwidth]{figure/rate-delta-indoor.pdf}
% 		\caption{Office.}
% 		\label{fig:rate-delta-indoor}
% 	\end{subfigure}
% 	\caption{Wi-Fi transmission rate change ratio in different environments.}
% 	\label{fig:moti-rate-delta}
% \end{figure}

%\nosection{Findings.}
\paragraph{An Example of Extended Packet Delivery Time.}
\figref{fig:moti-example} shows how a single packet's delivery time can extend to 75.9ms, 
orders of magnitude longer than expected. This significant extension stems from two factors. 
First, after each collision, the PPDU requires a retransmission attempt, 
and our example packet needs multiple retransmissions. 
Second, during each retransmission attempt, 
the contention interval becomes severely extended. 
While the doubled contention window only increases from CW=15 (max of $135\mu s$) to CW=31 (max of $279\mu s$), 
the actual contention intervals stretch to 43.5ms and 25.5ms because other devices (green triangles) repeatedly gain channel access during the countdown process. 
Each time another device transmits, our PPDU must freeze its countdown. 
Through this combination of multiple retransmission attempts and extended contention intervals, 
what should be a quick packet delivery becomes a 75.9ms process.

\paragraph{Statistical Analysis of Contention Intervals.}
To quantify this phenomenon, we analyzed the distribution of both PHY transmission times and contention intervals across all PPDUs (\figref{fig:delay-component-indoor}). PHY transmission time---the actual time spent transmitting data over the air as shown in \figref{fig:link-latency-decompose}---remains predictably brief ($<$ 5ms at 99.99th percentile) due to fixed Wi-Fi hardware constraints. In stark contrast, contention intervals, representing time spent competing for channel access, exhibit alarming variability. While their median stays below 1ms, the tail extends dramatically, exceeding 200ms at the 99.99th percentile. This means Wi-Fi devices spend orders of magnitude more time competing for transmission opportunities than actually transmitting data. 

\paragraph{Takeaway.}
Our real-world measurements validate that packet delivery drought stems from both retransmission attempts and extended contention intervals, with devices spending up to 200ms competing for channel access compared to just 5ms for actual data transmission.

\section{MAR-Driven CW Control Algorithm}
The detailed pseudo code of the MAR-driven CW control algorithm is shown in Alg.~\ref{alg:algo}.

\section{Target MAR Analysis}\label{appendix:design-theory}
The target microscopic access rate $MAR_{tar}$ plays an essential role in \sys's stable-state control policy. Here, we analyze the impact of $MAR_{tar}$ and discuss the criteria for selecting its optimal value.
\subsection{Inverse Proportion}
We first analyze the relationship between the $CW$ value and the microscopic access rate $MAR$. For a transmitter $i$ with contention window $CW_i$, the probability $\tau_i$ of attempting a transmission at any given transmission chance (highlighted in red in \figref{fig:tx-opportunity}) is the probability that its random backoff timer reaches zero at that moment:
\begin{align} 
\tau_i = \frac{CW_i}{\sum_{k=1}^{CW_i}k} = \frac{2}{CW_i + 1} 
\end{align}
In the stable state, the contention windows of all transmitters converge to the same value $CW$ (i.e., $\tau_i = \tau = \frac{2}{CW + 1}$). With $N$ transmitters, the probabilities for a transmission chance being idle, occupied by a successful transmission, or resulting in a collision are as follows:
\begin{align} 
    P_{i} = (1 - \tau)^N, \, P_{s} = N\tau(1 - \tau)^{N-1}, \, P_c = 1 - P_i - P_s \label{eqn:p}
\end{align}
The microscopic access rate $MAR$ represents the probability that a transmission chance is used:
\begin{align} 
    MAR = 1 - P_i = 1 - (1 - \tau)^N \stackrel{\tau \ll 1}{\approx} N\tau = \frac{2N}{CW + 1}\label{eqn:reverse-proportion}
\end{align}
Since $CW$ values are typically much larger than 1, we apply a first-order approximation in \eqnref{eqn:reverse-proportion}. This shows that, in the stable state, the microscopic access rate $MAR$ is inversely proportional to the converged contention window value $CW$.

\subsection{Robustness}
Here, we discuss the selection of the target MAR. In the stable state, the micro-level bandwidth fairness is achieved as all transmitters' $CW$ values converge to the same. Therefore, our target MAR aims to maximize the overall transmission bandwidth. Suppose the average PPDU size is $S$, the average time of a successful transmission, collided transmission, and slot time is $T_t$, $T_c$, and $T_s$, respectively. Note that \sys's design principle and control logic do not rely on these assumptions. Similar to the derivation in \cite{heusse05-idlesense}, the overall throughput can be presented as:
\begin{align}
    Thp = \frac{P_tS}{P_sT_t + P_cT_c + P_iT_s} = \frac{S}{T_t + \frac{(1 - P_i - P_s)\eta + P_i}{P_s} \cdot T_s}
\end{align}
where $\eta = T_c / T_s$. To maximize $Thp$, combined with \eqnref{eqn:p} and \eqnref{eqn:reverse-proportion}, we only have to minimize the cost function:
\begin{align}
    \mathcal{L}(MAR) &= \frac{(1 - P_i - P_s)\eta + P_i}{P_s} \nonumber \\
    &= \frac{N - MAR}{N} \cdot \frac{(\eta - 1)MAR + 1}{MAR(1-MAR)} \label{eqn:cost-function}
\end{align}
The optimal value $MAR_{opt}$ is determined by $N$ and $\eta$. However, since $MAR < 1$ and $N \ge 2$, $MAR$ has a negligible effect on the first term $\frac{N - MAR}{N}$. As a result, $MAR_{opt}$ is almost independent of the number of transmitters $N$. Next, consider the second term $\frac{(\eta - 1)MAR + 1}{MAR(1-MAR)}$. Taking the derivative of this term and setting it equal to zero yields:
\begin{align}
    MAR_{opt} = \frac{1}{\sqrt{\eta} + 1}
\end{align}

Since $\eta$ represents the number of time slots occupied by a collided transmission, its average value depends on the PHY transmission time of each PPDU, which in turn is determined by the PPDU size and PHY transmission rate. In the 802.11ax (Wi-Fi 6) standard, $\eta$ can range from 20 to over 500.
More importantly, although the optimal value of $MAR$ is primarily determined by $\eta$, the cost function $\mathcal{L}(MAR)$ is relatively insensitive to changes in $MAR$. We illustrate this by plotting $\mathcal{L}(MAR)$ against different values of $MAR$ and $\eta$ in a heatmap (\figref{fig:eta-mar-L}) with increasing $N$ values. As $MAR$ deviates from $MAR_{opt}$, the change in $\mathcal{L}(MAR)$ is minimal, and such pattern does not change with the increase of $N$ value. Therefore, \sys is robust to the selection of $MAR_{tar}$, as long as it remains within a "safe zone" ($\pm 0.1$) around $MAR_{opt}$. As a result, from \figref{fig:eta-mar-L}, we set the default value of $MAR_{tar}$ to be 0.1.

\section{Coexistence with IEEE 802.11 Standard Contention Control}\label{append:coexist}
\begin{table*}[htb]
\resizebox{\linewidth}{!}{
\begin{tabular}{ccccc}
\Xhline{2\arrayrulewidth}
                                    & $MAR_{tar} = 0.1$ / IEEE & $MAR_{tar} = 0.25$ / IEEE & $MAR_{tar} = 0.35$ / IEEE & $MAR_{tar} = 0.5$ / IEEE \\
\Xhline{2\arrayrulewidth}
\textbf{Avg. MAC Throughput (Mbps)} & 2.2 / 94.1             & 21.8 / 60.0             & 28.1 / 52.5             & 32.0 / 43.9            \\
\textbf{50th PPDU TX Delay (ms)}    & 224.8 / 4.6            & 21.3 / 6.5              & 16.1 / 7.2              & 13.7 / 7.9             \\
\textbf{95th PPDU TX Delay (ms)}    & 491.7 / 11.3           & 48.6 / 21.6             & 38.9 / 25.1             & 38.9 / 25.1            \\
\textbf{99th PPDU TX Delay (ms)}    & 634.1 / 17.9           & 63.9 / 38.8             & 52.9 / 48.5             & 52.9 / 48.5            \\
\textbf{99.9th PPDU TX Delay (ms)}  & 888.2 / 34.9           & 88.8 / 85.7             & 72.2 / 112.6            & 72.2 / 112.6      \\
\Xhline{2\arrayrulewidth}
\end{tabular}
}
\caption{The performance of \sys coexisting with IEEE 802.11 standard contention control policy.}
\label{tab:coexist}
\end{table*}

Similar to the experimental setup in \secref{sec:eval-wild}, we deploy four AP-STA pairs, with two pairs running \sys and the other two pairs using the IEEE 802.11 standard contention control policy. The APs saturate the wireless link by sending \texttt{iperf} traffic to the STAs. As shown in \tabref{tab:coexist}, with $MAR_{tar}$ increasing from 0.1 to 0.5, \sys becomes more competitive to the standard control policy and gains more MAC throughput with lower PPDU transmission delay. Therefore, \sys can be configured with higher $MAR_{tar}$ values when coexisting with the standard policy, while still ensuring convergence and fairness when all APs implement \sys.

\section{Influence of Hidden Terminal}\label{append:hidden-terminal}
Since \sys relies on the consensus signal from the same channel, it may be affected by the Hidden Terminal Problem \cite{hidden-terminal}. To demonstrate this impact, similar to the experimental setup in \secref{sec:eval-wild}, but we deploy the AP-STA pairs in three rooms arranged in a row. Transmitters at both ends cannot hear each other, acting as hidden terminals, while transmitters in the middle can hear traffic from both ends, acting as exposed terminals. We show the PPDU transmission delay distribution for hidden and exposed terminals in \figref{fig:micro-hidden}. When RTS/CTS is disabled, both \sys and the IEEE 802.11 standard contention control policy result in increased tail latency for exposed terminals, as they undergo more intensive contention. However, with RTS/CTS enabled, because \sys counts CTS signals in opportunity utilization rate calculation, \sys shows much smaller differences in delay distribution between exposed and hidden terminals. These results demonstrate that \sys is compatible with the widely deployed RTS/CTS mechanism and is robust to the Hidden Terminal Problem when RTS/CTS is enabled.

\section{\sys Contention Window Control Algorithm}
\label{sec:Contention Window Control Algorithm}

We show the pseudo-code of \sys contention window control algorithm in \algoref{alg:algo}.

\noindent
\begin{algorithm}[t]
	\caption{\sys contention window control.}
	\label{alg:algo}
        \textbf{Parameters}
        
        $N_{obs}, MAR_{tar}, MAR_{max}$ \Comment{Default 300, 0.1, 0.35}

        $CW_{min}, CW_{max}$ \Comment{Default 15, 1023}

        $M_{inc}, M_{dec}, A_{inc}, A_{fail}$ \Comment{Default 500, 0.95, 15, 5}

        \
        
	\textbf{Initialization}

        $N_{idle} \leftarrow 0,$
        $N_{tx} \leftarrow 0,$
        $first\_rtx \leftarrow True,$
        
        $CW \leftarrow CW_{min},$
        $CW_{fail} \leftarrow CW$

        \

        \textbf{Func} OnNewWNICState($state$, $duration$)

		\If{ $state$ = IDLE }{
			$N_{idle} \leftarrow N_{idle} + duration / slot\_time$
		}
		\If {$state$ = BUSY} {
                $N_{tx} \leftarrow N_{tx} + 1$
            }

        \
        
        \textbf{Func} OnACK() \Comment{Stable Control Policy}

            $CW \leftarrow CW_{fail}$ \Comment{Restore the CW at previous failure}
            
            \If {$N_{idle} + N_{tx} < N_{obs}$} {
                \textbf{return} \Comment{No enough samples}
            }
            $MAR \leftarrow N_{tx} / (N_{tx} + N_{idle})$

            \If {$MAR > MAR_{tar}$} {

                    $CW \leftarrow CW + CW \cdot \max\{0, MAR - MAR_{max}\}  + M_{inc}(\min \{MAR, MAR_{max}\} - MAR_{tar}) + A_{inc}$

            }
            \Else {
                ${\scriptsize CW \leftarrow \min\{M_{dec} - \frac{(1\!-\!M_{dec})(CW\!-\!CW_{min})}{CW_{max}\!-\!CW_{min}}, \frac{2MAR}{MAR_{tar}\!+\!MAR} \} \times CW}
$

            }

            $N_{idle} \leftarrow 0,$
            $N_{tx} \leftarrow 0,$
            
            $CW_{fail} \leftarrow CW,$
            $first\_rtx \leftarrow True$

        \

        \textbf{Func} OnACKFailure() \Comment{Fast Recovery from Collision}

        % $CW_{fail} \leftarrow CW + A_{fail}$ \Comment{Increase and store CW}
        % $U_{obs} \leftarrow busy\_slots / (busy\_slots + idle\_slots)$

        \If {$first\_rtx$} {
            $CW_{fail} \leftarrow CW + A_{fail}$ \Comment{Increase and store CW}
            
            $CW \leftarrow CW_{fail} / 2$ \Comment{Accelerated retransmission}
            
            $first\_rtx \leftarrow False$ \Comment{Only accelerate once}
        }
        
        % \If {$U_{obs} < U_{max}$} {
        %     $CW \leftarrow CW_{fail} / 2$ \Comment{Accelerated retransmission}
        % }
\end{algorithm}

\section{Observation Interval Analysis}\label{appendix:observation-interval}

Assume that all Wi-Fi devices in the current Wi-Fi environment are collaboratively working to stabilize the $MAR$ around the target value $MAR_{\mathrm{tar}}$. We assume that the state of the channel being busy is represented by 1, and the state of the channel being idle is represented by 0. These busy or idle states of the channel over a period of time form an approximate i.i.d.\ sequence of Bernoulli random variables $X_i$, with success probability $MAR_{\mathrm{tar}} = 0.15$. The sample mean over $N_{\mathrm{obs}}$ observations is:

\[
\overline{X}_{N_{\mathrm{obs}}} = \frac{1}{N_{\mathrm{obs}}} \sum_{i=1}^{N_{\mathrm{obs}}} X_i.
\]

For $N_{\mathrm{obs}} = 300$, the standard error (SE) of $\overline{X}_{300}$ is:

\[
\text{SE}(\overline{X}_{300}) = \sqrt{\frac{0.15 \times 0.85}{300}} \approx 0.0206.
\]

By the Chernoff bound for binomial distributions, the probability of deviation exceeding $\delta$ satisfies:

\[
\mathbb{P}(|\overline{X}_{N_{\mathrm{obs}}} - MAR_{\mathrm{tar}}| \geq \delta) \leq 2\exp\left(-\frac{N_{\mathrm{obs}} \delta^2}{3 MAR_{\mathrm{tar}}(1 - MAR_{\mathrm{tar}})}\right).
\]

For $N_{\mathrm{obs}} = 300$ and $\delta = 0.02$:

\[
\mathbb{P}(|\overline{X}_{300} - 0.15| \geq 0.02) \leq 2e^{-0.314} \approx 1.462\%.
\]

This confirms that the estimation error remains negligible with high probability. So $N_{obs} = 300$ is Sufficient.

\section{Collision Probability in Wi-Fi Networks}
\label{sec:Collision Probability in WiFi}
\begin{figure}[t]
    % \centering
    \includegraphics[width=0.8\linewidth]{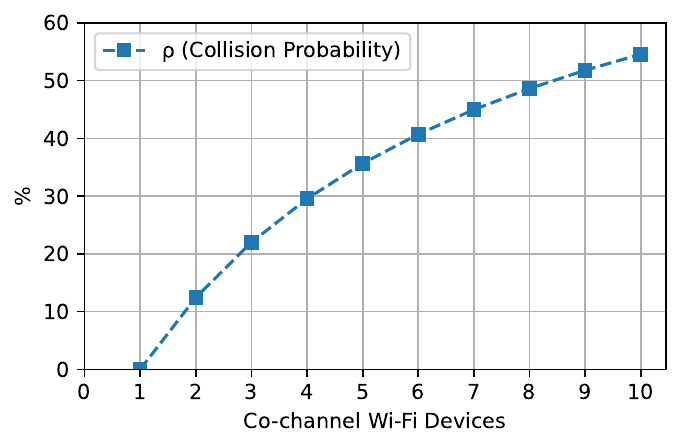}
    \caption{Collision probability vs. number of co-channel Wi-Fi devices (when the transmission queue remains non-empty).}
    \label{fig:Collision_Probability}
\end{figure}

Consider a scenario with $N$ Wi-Fi devices, where each device has a transmission probability of $\tau$ in an arbitrary time slot and the transmission queue of each device remains non-empty. The collision probability, denoted as $\rho$, can be expressed as:
\begin{equation}
\rho = 1 - (1 - \tau)^{N-1}.\label{eqn:rho}
\end{equation}
Assume all Wi-Fi devices operate in the BE (Best Effort) queue, where the contention window ($CW$) doubles from $CW_{\text{min}}$ to $CW_{\text{max}}$ after each retransmission, up to a maximum of $r$ retransmissions.
Suppose $x$ transmissions use $CW_{\text{min}}$, and $x\rho^i$ transmissions use $CW_{\text{min}}2^i$ for $0 \leq i \leq r$. The probability of transmitting with $CW_{\text{min}}2^i$ is then given by:
\begin{equation}
P_i = \frac{x\rho^i}{\sum_{j=0}^{r} x\rho^j} = \frac{\rho^i}{\sum_{j=0}^{r} \rho^j}.\label{eqn:P_i}
\end{equation}
Thus, the transmission probability $\tau$ can be derived as:
\begin{equation}
\tau = \sum_{i=0}^{r} \frac{2P_i}{CW_{\text{min}}2^i}.\label{eqn:tau}
\end{equation}
By solving Eqn.~\ref{eqn:rho}, \eqnref{eqn:P_i} and \eqnref{eqn:tau} simultaneously, the solution for $\rho$ is obtained numerically using the bisection method within the range $(0, 1)$, ensuring convergence to the unique solution for each $N$. 
\figref{fig:Collision_Probability} illustrates the variation in collision probability as the number of co-channel Wi-Fi devices increases. The analysis assumes that all Wi-Fi devices operate under the BE queue and maintain continuously non-empty transmission queues. The results indicate that when the number of co-channel Wi-Fi devices reaches 10, the collision probability exceeds 50\%, highlighting the significant impact of device density on network performance.

\section{When MAR is Fixed, the Collision Probability is Constrained Below MAR}
\label{sec:Collision Probability upper bound}

Consider a scenario where $N$ WiFi devices continuously transmit data packets. Assuming that the contention window (CW) value for each device is $\omega - 1$, the attempt probability $\tau$ (i.e., the probability that any given WiFi device completes its random backoff and begins transmission at any slot-time \cite{Bianchi2000} can be expressed as:

\begin{equation}
\tau = \frac{\omega}{\frac{\omega^2}{2}} = \frac{2}{\omega}
\end{equation}

%Here, the denominator represents the sum of all possible backoff values from 0 to $\omega$, while the numerator accounts for the total occurrences of countdown completions from all possible backoff values $b$.

The probability of a collision occurring when a WiFi device attempts transmission, denoted as $\rho$, is given by \eqnref{eqn:rho}.
The MAR is defined as the probability that the channel is in a busy state, which corresponds to at least one WiFi device transmitting. This can be expressed as:

\begin{equation}
MAR = 1 - (1 - \tau)^N
\end{equation}

Given that $\tau = \frac{2}{\omega} \in (0,1)$ and is relatively close to zero, it follows that:

\begin{equation}
MAR = 1 - (1 - \tau)^N > 1 - (1 - \tau)^{N-1} = \rho
\end{equation}

%\begin{equation}
%\rho < MAR
%\end{equation}

This implies that, when MAR is determined, the collision probability remains stable at a level below MAR.